\documentclass[11pt, epsfig]{article}
\usepackage{epsfig, amsmath, amssymb, amsthm, times}
\usepackage{graphicx}
\def\theequation{\arabic{section}.\arabic{equation}}
\renewcommand{\theequation}{\thesection.\arabic{equation}}

\newtheorem {thm}{Theorem}[section]
\newtheorem {lem}[thm]{Lemma}

\theoremstyle{defintion}

\theoremstyle{remark}
\newtheorem{rem}[thm]{Remark}

\theoremstyle{example}
\newtheorem{ex}[thm]{Example}

\theoremstyle{assumption}
\newtheorem{assume}[thm]{Assumption}

\def\pf{{\it Proof.\;}}
\def\var{\mathrm{var}~}

\def\E{{\mathbb E~}}
\def\P{{\mathbb P~}}
\def\R{{\mathbb R}}
\def\N{{\mathbb N}}

\def\lbl{\label}
\def\be{\begin{equation}}
\def\ee{\end{equation}}
\def\p{\partial}
\def\qed{\square}
\def\tr{\mathrm{Tr}~}
\def\row{\mathrm{Row}}

\def\t{\mathsf{T}}
\def\tl{\tilde}

\title{The geometry of spontaneous spiking in neuronal networks
}
\author{Georgi S. Medvedev\thanks{
Department of Mathematics, Drexel University, 3141 Chestnut Street,
Philadelphia, PA 19104, {\tt medvedev@drexel.edu} 
}\;
  and  Svitlana Zhuravytska\thanks{Bank of America, 
1200 N. King St., Wilmington, DE 19801,
{\tt svitlana.zhuravytska@bankofamerica.com} 
}
}
\begin{document}
\maketitle
\begin{abstract} 
The mathematical theory of pattern formation in electrically coupled networks
of excitable neurons forced by small noise is presented in this work.
Using the Freidlin-Wentzell large deviation theory for randomly perturbed dynamical
systems and the elements of the algebraic graph theory, we identify and analyze 
the main regimes in the network dynamics in terms of the key control parameters:
excitability, coupling strength, and network topology. The analysis reveals the 
geometry of spontaneous dynamics in electrically coupled network.
Specifically, we show that
the location of the minima of a certain continuous function on the surface of the
unit $n-$cube encodes the most likely activity patterns generated by the network.
By studying  how the minima of this function evolve under the variation of the 
coupling strength, we describe the principal transformations in the network 
dynamics. The minimization problem is also used for the quantitative description 
of the main dynamical regimes and transitions between them. In particular, 
for the weak and strong coupling regimes, we 
present asymptotic formulae for the network activity rate as a function of the coupling
strength and the degree of the network. The variational analysis is complemented 
by the stability analysis of the synchronous state in the strong coupling regime.
The stability estimates reveal the contribution 
of the network connectivity and the properties of the cycle subspace associated
with the graph of the network to its synchronization properties.
This work is motivated by the experimental and modeling studies of the ensemble
of  neurons in the  Locus Coeruleus, a nucleus in the brainstem involved in 
the regulation of cognitive performance and behavior.  
\end{abstract}

\section{Introduction}\lbl{intro}
\setcounter{equation}{0}
Direct electrical coupling through gap-junctions is a common way of communication 
between neurons, as well as between cells of the heart, pancreas, and other physiological 
systems \cite{CL04}. Electrical synapses are important for synchronization of the
network activity, wave propagation, and  pattern formation in neuronal networks.
A prominent example of a gap-junctionally coupled network, whose dynamics is thought 
to be important for cognitive processing, is a group of neurons in the 
Locus Coeruleus (LC), a nucleus in the brainstem \cite{AJC05, BW03, SA09}. 
Electrophysiological studies of the animals performing a visual discrimination 
test show that the rate and the pattern of activity of the LC network correlate with 
the cognitive performance \cite{UCS99}.  Specifically, the periods of the high spontaneous
activity correspond to the periods of poor performance, whereas the periods of low
synchronized activity coincide with good performance. Based on the physiological
properties of the LC network, it was proposed that the 
transitions between the periods of high and low network activity are due to the 
variations in the strength of coupling between the LC neurons \cite{UCS99}. This
hypothesis motivates the following dynamical problem: to study how the dynamics 
of  electrically coupled networks depends on the coupling strength. This question
is the focus of the present work.
 
The dynamics of an electrically coupled network depends on the properties of the 
attractors of the local dynamical systems and the interactions between 
them. Following \cite{UCS99}, we assume that the individual neurons in the LC network are
spontaneously active. Specifically, we model them with excitable dynamical systems
forced by small noise. We show that depending on the strength of electrical coupling,
there are three main regimes of the network dynamics: uncorrelated spontaneous firing 
(weak coupling), formation of clusters and waves (intermediate coupling), and 
synchrony (strong coupling). The qualitative features of these regimes are
independent from the details of the models of the individual neurons and
network topology. Using the center manifold
reduction \cite{CH82, GH} and the Freidlin-Wentzell large deviation theory
\cite{FW}, we derive a variational
problem, which provides a useful geometric interpretation for various patterns
of spontaneous activity.
Specifically, we show that
the location of the minima of a certain continuous function on the surface of the
unit $n-$cube encodes the most likely activity patterns generated by the network.
By studying  the evolution of the minima of this function under the variation of 
the control parameter (coupling strength), we identify the principal transformations 
in the network dynamics. The minimization problem is also used for the quantitative 
description  of the main dynamical regimes and transitions between them. In particular, 
for the weak and strong coupling regimes, we present asymptotic formulae for the 
activity rate as a function of the coupling strength and the degree of the network. 
The variational analysis is complemented by the stability analysis of the synchronous 
state in the strong coupling regime. 
In analyzing various aspects of the network dynamics,
we pay special attention to the role of the structural properties of the network 
in shaping its dynamics. We show  that in weakly coupled networks, only very rough 
structural properties of the underlying graph matter, whereas in the strong
coupling regime, the finer features, such as the algebraic connectivity and 
the properties of the cycle subspace associated with the graph of the network, 
become important.
Therefore, this paper presents a comprehensive analysis of electrically
coupled networks of excitable cells in the presence of noise. It complements
the existing studies of related deterministic networks of electrically 
coupled oscillators (see, e.g., \cite{Coombes, GH07, LR03, MK01} and references therein).
 
The outline of the paper is as follows.
In Section~\ref{themodel}, we formulate the biophysical model of the LC network. 
Section~\ref{s3} presents numerical experiments elucidating the principal 
features of the network dynamics. 
In Section~\ref{analysis}, we reformulate the problem in terms of the 
bifurcation properties of the local dynamical systems and the properties
of the linear coupling operator. We then introduce the variational problem,
whose analysis explains the main dynamical regimes of the coupled system.
In Section~\ref{another}, we analyze the stability of the synchronous 
dynamics in the strong coupling regime, using fast-slow decomposition.
The results of this work are summarized  in Section~\ref{discuss}. 

\section{The model}\lbl{themodel}
\subsection{The single cell model}\lbl{sec1.1}
According to the dynamical mechanism underlying action potential generation, 
conductance-based models of neurons  are divided into Type I and Type II
classes \cite{RE89}. The former assumes that the model is near the saddle-node
bifurcation, while the latter is based on the Andronov-Hopf bifurcation.
Electrophysiological recordings of the LC neurons  exhibit features that
are consistent with the Type I excitability.  The existing biophysical 
models of LC neurons use Type I action potential generating
mechanism \cite{ACV02, BMH04}. In accord with these experimental and modeling 
studies, we use a generic Type I conductance-based model to  simulate 
the dynamics of the individual LC neuron
\begin{eqnarray}\lbl{1.1}
C\dot v &=& -I_{ion}(v, n)+\sigma \dot w,\\
\lbl{1.2}
\dot n &=& {n_\infty(v)-n\over \tau(v)}.
\end{eqnarray}
Here, dynamical variables $v(t)$ and $n(t)$ are the membrane
potential and the activation of the potassium current, $I_K$, respectively.
$C$ stands for the membrane capacitance. 
The ionic currents $I_{ion}(v,n)$ are modeled using the 
Hodgkin-Huxley formalism (see Appendix for the definitions
of the functions and parameter values used in (\ref{1.1}) and (\ref{1.2})).
A small Gaussian white 
noise is added to the right hand side of (\ref{1.1}) to simulate 
random synaptic input and other possible fluctuations affecting system's dynamics. 
Without noise ($\sigma=0$), the system is in the excitable regime.
For $\sigma>0$, it exhibits spontaneous spiking. The frequency of the 
spontaneous firing depends on the proximity of the deterministic system
to the saddle-node bifurcation and on the noise intensity.
A typical trajectory of (\ref{1.1}) and (\ref{1.2}) stays
in a small neighborhood of the stable equilibrium most of the time (Fig.~\ref{f.1}a). 
Occasionally, it leaves the vicinity of the fixed point to  make a large excursion 
in the phase plane and then returns to the neighborhood of the steady state (Fig.~\ref{f.1}a).
These dynamics generate a train of random spikes in the voltage time series
(Fig.~\ref{f.1}b).
\begin{figure}
{\bf a}\includegraphics[height=2.0in,width=2.5in]{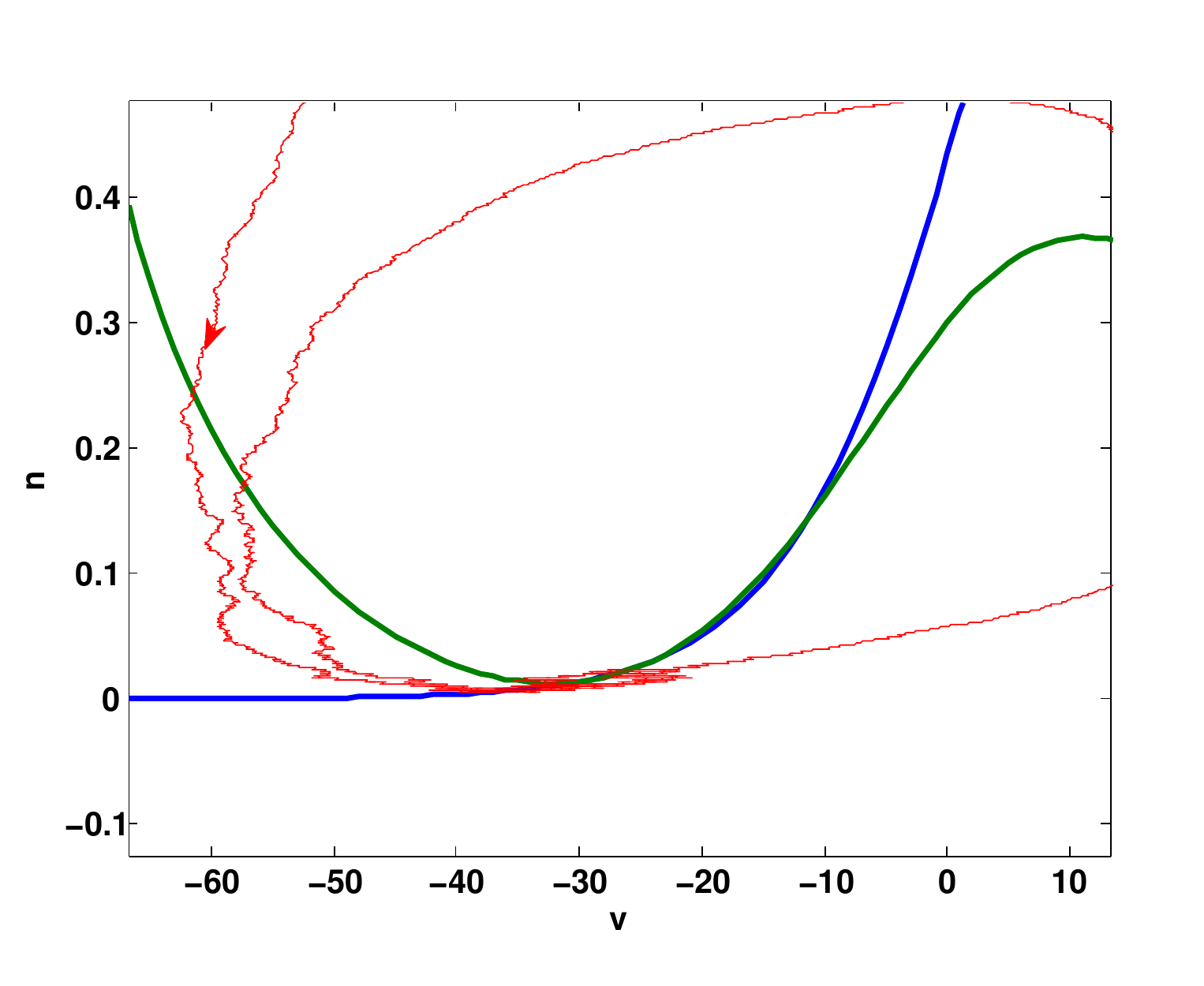} \quad
{\bf b}\includegraphics[height=2.0in,width=3.0in]{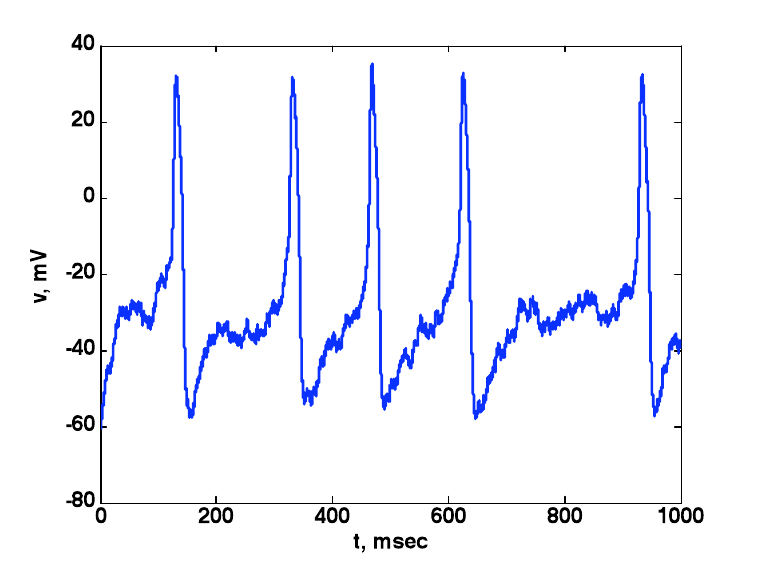}
\caption{ {\bf a}) The phase plane for (\ref{1.1}), (\ref{1.2}): nullclines
plotted for the deterministic model ($\sigma=0$) and a trajectory
of the randomly perturbed system ($\sigma>0$). 
The trajectory spends most time in a small neighborhood of the 
stable fixed point. Occasionally, it leaves the basin of attraction of the
fixed point to generate a spike. {\bf b}) The voltage timeseries, $v(t)$,
corresponding to spontaneous dynamics shown in plot \textbf{a}.
}
\lbl{f.1}
\end{figure}

In neuroscience, the (average) firing rate provides a convenient measure of activity
of neural cells and neuronal populations. It is important to know how the firing
rate depends on the parameters of the model. In this paper, we study the factors determining
the rate of firing in electrically coupled network of neurons.
However, before setting out to study the network dynamics, it is instructive to discuss 
the behavior of the single neuron model first. To this end, we use the center-manifold 
reduction to approximate (\ref{1.1}) and (\ref{1.2})
by a $1D$ system:
\be\lbl{1d}
\dot z =-U^\prime(z)+ \tilde\sigma \dot w_t, \;
U(z)=\mu z-{1\over 3} z^3+{2\over 3}\mu^{3/2},
\ee
where $z(t)$ is the rescaled projection of $\left(v(t),n(t)\right)$ onto a $1D$ slow
manifold, $\mu>0$ is the distance
to the saddle-node bifurcation, and $\tilde\sigma>0$ is the noise intensity after rescaling.
We postpone the details of the center-manifold reduction until we analyze a more general network
model in \S\ref{center}. 

The time between two successive spikes in voltage time series corresponds to the first time 
the trajectory of (\ref{1d}) with initial condition $z(0)=z_0<\sqrt{\mu}$ overcomes 
potential barrier $U(\sqrt{\mu})-U(-\sqrt{\mu})$. 
The large deviation estimates (cf. \cite{FW}) yield the logarithmic
asymptotics of the first crossing time $\tau$ 
\be\lbl{arhenius}
\lim_{\tilde\sigma\to 0} \tilde\sigma^2\ln\E_{z_0}\tau=
2U(\sqrt\mu)={4\mu^{3/2}\over 3}\; \Rightarrow\; 
\E_{z_0}\tau \asymp \exp\left\{{ 4\mu^{3/2}\over 3\tilde\sigma^{2}}\right\},
\ee
where
$\E_{z_0}$ stands for the expected value with respect to the 
probability generated by the random process $z(t)$
with initial condition $z(0)=z_0$. Throughout this paper, we use $\asymp$
to denote logarithmic asymptotics. It is also known that the first exit time $\tau$ is 
distributed exponentially as shown in Fig.~\ref{f.2} (cf. \cite{Day83}).

Equation (\ref{arhenius}) implies that the statistics of spontaneous spiking of a single cell
is determined by the distance of the neuronal model (\ref{1.1}) and (\ref{1.2})
to the saddle-node bifurcation and the intensity of noise. Below we show that, 
in addition to these two parameters, the strength and topology of coupling are important 
factors determining the firing rate of the coupled population. 

\begin{figure}
\begin{center}
\includegraphics[height=2.0in,width=2.2in]{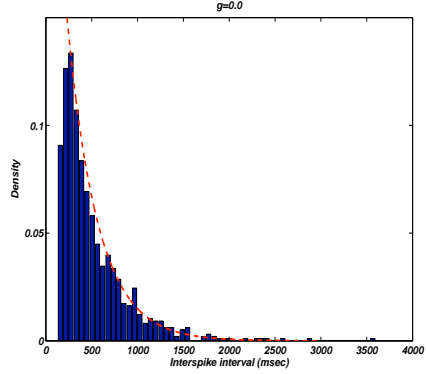}
\end{center}
\caption{ The numerical approximation of the density of the time between
the successive spikes in the voltage timeseries $v(t)$, obtained by
integration (\ref{1.1}) and (\ref{1.2}). The interspike intervals are 
distributed approximately exponentially.
}
\lbl{f.2}
\end{figure}

\subsection{The electrically coupled network}
\lbl{s1.2}
The network model includes $n$ cells, whose intrinsic dynamics is defined by 
(\ref{1.1}) and (\ref{1.2}), coupled by gap-junctions.
The gap-junctional current that Cell $i$ receives from the other cells in the network 
is given by
\be\lbl{1.3}
I_c^{(i)}=g\sum_{i=1}^n a_{ij} \left(v^{(j)}-v^{(i)}\right),
\ee
where $g\ge 0$ is the gap-junction conductance
and 
$$
a_{ij}=\left\{\begin{array}{cc}
1, & \mbox{Cell~$i$ and Cell~$j$ are connected},\\
0, & \mbox{otherwise}.
\end{array}
\right.\;\;\; a_{ii}=0,\; (i,j)\in [n]^2.
$$ 
Adjacency matrix $A=(a_{ij})\in\R^{n\times n}$  defines the network connectivity.
By adding the coupling current to the right hand side of the voltage equation (\ref{1.1})
and combining the equations for all neurons in the network, 
we arrive at the following model 
\begin{eqnarray} \lbl{1.4}
C\dot v^{(i)} &=& -I_{ion}(v^{(i)}, n^{(i)})+
g\sum_{i=1}^n a_{ij} \left(v^{(j)}-v^{(i)}\right)+\sigma \dot w^{(i)},\\
\lbl{1.5}
\dot n^{(i)} &=& {n_\infty(v^{(i)})-n^{(i)} \over\tau(v^{(i)})},
\end{eqnarray}
where $ w^{(i)}$ are $n$ independent copies of the standard Brownian motion.
\begin{figure}
\begin{center}
{\bf a}\includegraphics[width=0.2\textwidth]{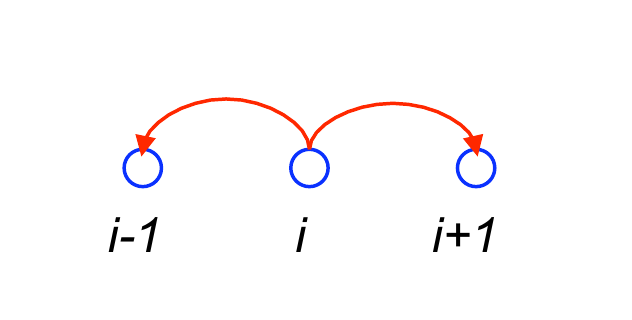}\qquad\qquad
{\bf b}\includegraphics[width=0.2\textwidth]{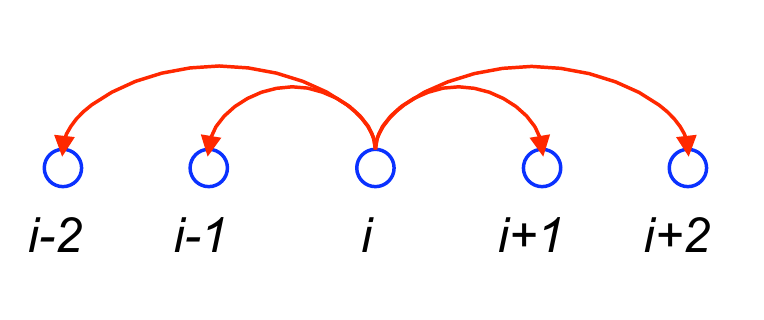}\qquad\qquad
{\bf c}\includegraphics[width=0.2\textwidth]{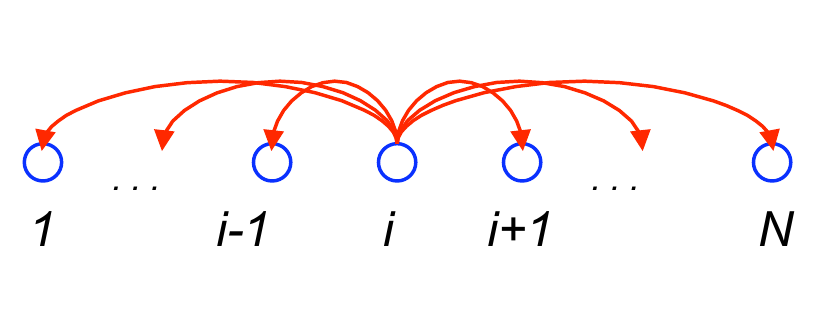} \\
\end{center}
\caption{
Examples of coupling schemes: (a) nearest neighbor
(cf. Example~\ref{ex.1}); (b) $2-$nearest neighbor (cf. Example~\ref{ex.2});  
(c) all-to-all (cf. Example~\ref{ex.3}).
}
\lbl{f.3}
\end{figure}

The network topology is an important parameter of the model (\ref{1.4}) and (\ref{1.5}).
The following terminology and constructions from the algebraic graph theory  \cite{Biggs}
will be useful for studying the role of the network structure in shaping its dynamics.
Let $G=(V(G),E(G))$ denote the  graph of interactions between the cells in the network.
Here, $V(G)=\{v_1, v_2, \dots, v_n \}$ and $E(G)=\{e_1, e_2, \dots,e_m\}$
denote the sets of vertices (i.e., cells) and  edges 
(i.e., the pairs of connected cells), respectively.
Throughout this paper, we assume that $G$ is a connected graph.
For each edge $e_j=(v_{j_1},v_{j_2})\in V(G)\times V(G)$, we 
declare one of the vertices $v_{j_1}, v_{j_2}$ to be the positive end (head)
of $e_j$, and the other to be the negative end (tail). Thus, we assign an 
orientation to each edge from its tail to its head.  
The coboundary matrix of $G$ 
is defined as follows (cf. \cite{Biggs}) 
\be\lbl{incidence}
H=(h_{ij})\in \R^{m\times n},\quad
h_{ij}=\left\{ \begin{array}{cl}
1, & v_j\;\mbox{ is a positive end of}\; e_i,\\
-1, & v_j\;\mbox{ is a negative end of}\; e_i,\\
0, &\;\mbox{otherwise}.
\end{array}
\right.
\ee
Let $\tilde G=(V(\tilde G),E(\tilde G))\subset G$ be a spanning tree of $G$,
i.e., a connected subgraph of $G$ such 
that $|V(\tilde G)|=n$, and there are no cycles in $\tilde G$ \cite{Biggs}. 
Without loss of generality, we assume that 
\be\lbl{spanning}
E(\tilde G)=\{e_1, e_2, \dots, e_{n-1}\}.
\ee
Denote the coboundary matrix of $\tilde G$ by $\tilde H$.

Matrix 
\be\lbl{Lap-1}
L=H^\t H
\ee
is called a graph Laplacian of $G$. The Laplacian is 
independent of the choice of orientation of edges that was used
in the definition  of $H$ \cite{Biggs}.
Alternatively, the Laplacian can be defined as
\be\lbl{def-Lap}
L=D-A, 
\ee
where 
$D=\mathrm{diag}\{\mathrm{deg}(v_1),\mathrm{deg}(v_2),\dots
\mathrm{deg}(v_n)\}$ is the degree map and $A$ is the adjacency matrix
of $G$. 
 
Let 
$$
\lambda_1(L)\le \lambda_2(L)\le\dots\le \lambda_{n}(L)
$$
denote the eigenvalues of $L$ arranged
in the increasing order counting the multiplicity.
The spectrum of the graph Laplacian captures many structural properties
of the network (cf. \cite{Biggs, Bollobas98, Chung97}). 
In particular, the first  eigenvalue of $L$, $\lambda_1(L)=0$,
is simple if and only if the graph is connected \cite{Fiedler73}. 
The second eigenvalue $\mathfrak{a}=\lambda_2(L)$ is called the algebraic 
connectivity of $G$, because it yields a lower bound for the edge and the vertex connectivity of 
$G$ \cite{Fiedler73}. 
The algebraic connectivity is important  for a variety
of combinatorial, probabilistic, and dynamical  aspects of the network analysis.
In particular, it is used in the studies of the graph expansion \cite{Hoory06}, 
random walks \cite{Bollobas98}, and synchronization of dynamical networks
\cite{Jost07, medvedev10b}.

Next, we introduce several examples of the network connectivity including
nearest neighbor arrays of varying degree and a pair of degree $4$ symmetric
and random graphs. These examples will be used
to illustrate the role of the network topology in pattern formation.
\begin{figure}\begin{center}
{\bf a}\includegraphics[height=2in,width=2in]{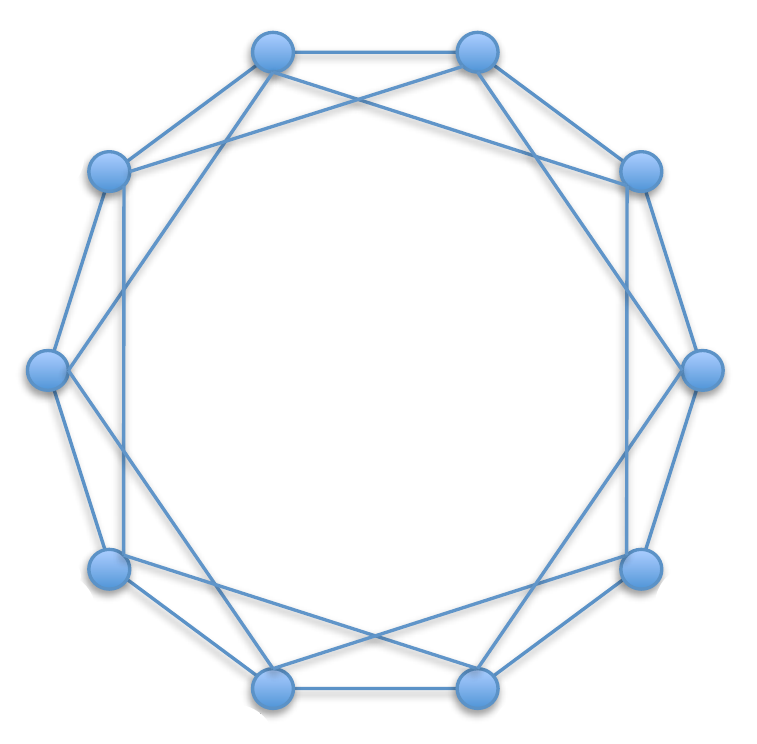} \hspace{0.8in}
{\bf b}\includegraphics[height=2in,width=2in]{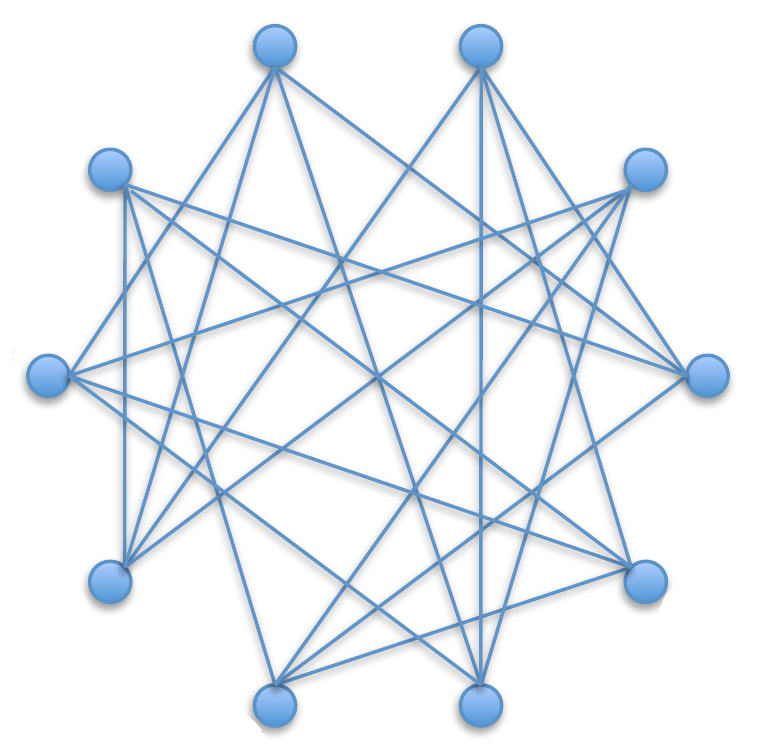}
\end{center}
\caption{ 
Regular versus random connectivity. Both graphs in (a) and (b) have degree $4$.
The graph in (a) is formed using regular coupling scheme, whereas edges of the
graph in (b) are generated using a random algorithm (cf. Example~\ref{ex.4}).
}
\lbl{f.3a}
\end{figure}

\begin{ex}\lbl{ex.1} 
The nearest-neighbor coupling scheme is an example of 
the local connectivity (Fig.~\ref{f.3}a). For simplicity, we consider a $1D$ array.
For higher dimensional lattices, the nearest neighbor coupling is defined similarly.
In this configuration,
each cell in the interior of the array is coupled
to two nearest neighbors. This leads to the following expression for the 
coupling current:
$$
I_c^{(j)}=g(v^{(j+1)}-v^{(j)})+g(v^{(j-1)}-v^{(j)}),\;
j=2,3,\dots,n-1.
$$
The coupling currents for the cells on the boundary
are given by
$$  
I_c^{(1)}=g(v^{(2)}-v^{(1)})\quad\mbox{and}\quad 
I_c^{(n)}=g(v^{(n-1)}-v^{(n)}).
$$
The corresponding graph Laplacian is 
\be\lbl{1.6}
L=\left(\begin{array}{cccccc}
1 &-1 & 0& \dots &0&0 \\
-1 & 2& -1 & \dots& 0&0\\
\dots&\dots&\dots&\dots&\dots&\dots\\
0 &0 & 0& \dots &-1 &1
\end{array}
\right).
\ee
\end{ex}
\begin{ex}\lbl{ex.2}
 The $k-$nearest neighbor coupling scheme is a natural generalization of the previous
example. Suppose 
 each cell is coupled to $k$ of its nearest neighbors from
 each side whenever they exist or as many as possible
 otherwise:
 \be\lbl{1.7}
 I_c^{(j)}=\sum_{i=1}^{\min\{k,n-j\} } g(v^{(j+i)}-v^{(j)})
 +\sum_{i=1}^{\min\{k,j\} } g(v^{(j-i)}-v^{(j)}),\;
 j=2,3,\dots,n-1,
 \ee
 where we use a customary convention that $\sum_{j=a}^b\dots=0$ if
 $b<a$. The coupling matrix can be easily derived from (\ref{1.7}).
 \end{ex}
\begin{ex}\lbl{ex.3}
The all-to-all coupling features global connectivity (Fig.~\ref{f.3}c):
\be\lbl{1.8}
I_c^{(j)}=g\sum_{i=1}^n (v^{(i)}-v^{(j)}),\;j=1,2,3,\dots,n.
\ee
The Laplacian in this case has the following form
\be\lbl{1.9}
L=\left(\begin{array}{cccccc}
n-1 &-1 & -1& \dots &-1&-1 \\
-1 & n-1& -1 & \dots& -1&-1\\
\dots&\dots&\dots&\dots&\dots&\dots\\
-1 &-1 & -1& \dots &-1 &n-1
\end{array}
\right).
\ee
\end{ex}
The graphs in the previous examples have different degrees:
ranging from $2$ in Example~\ref{ex.1} to $n-1$ in Example~\ref{ex.3}.
In addition to the degree of the graph, the pattern of connectivity itself
is important for the network dynamics. This motivates our
next example.
\begin{ex}\lbl{ex.4}
Consider a pair of degree $4$ graphs shown schematically
in Fig.~\ref{f.3a}. The graph in Fig.~\ref{f.3a}a has symmetric
connections. The edges of the graph in  Fig.~\ref{f.3a}b were 
selected randomly. Both graphs have the same number of nodes
and equal degrees.
\end{ex}
Graphs with random connections like the one in the last example
represent expanders, a class of graphs used in many important applications 
in mathematics, computer science and other branches of science and technology
(cf. \cite{Hoory06}). In Section~\ref{another}  we show that dynamical networks on expanders
have very good synchronization properties (see also \cite{medvedev10b}).

\begin{ex}\lbl{ex.5} Let $\{G_n\}$ be a family 
of graphs on $n$ vertices, with the following property:
\be\lbl{expander}
\lambda_2(G_n)\ge \alpha>0,\quad n\in\N.
\ee
Such graphs are called (spectral) expanders \cite{Hoory06, Sar04}.
There are known explicit constructions of expanders, including the 
celebrated Ramanujan graphs \cite{Margulis88, LPS88}. 
In addition, families of random graphs have good expansion properties. 
In particular, it is known that
\be\lbl{Friedman}
\mathrm{Prob}\left\{ \lambda_2(G_n)\ge d-2\sqrt{d-1}-\epsilon\right\}=1-o_n(1) \; \forall\epsilon>0,
\ee
where $G_n$ stands for the family of random
graphs of degree $d\ge 3$ and $n\gg 1$ \cite{Fri08}.
\end{ex}
\begin{figure}
\begin{center}
{\bf a}\includegraphics[height=1.8in,width=2.0in]{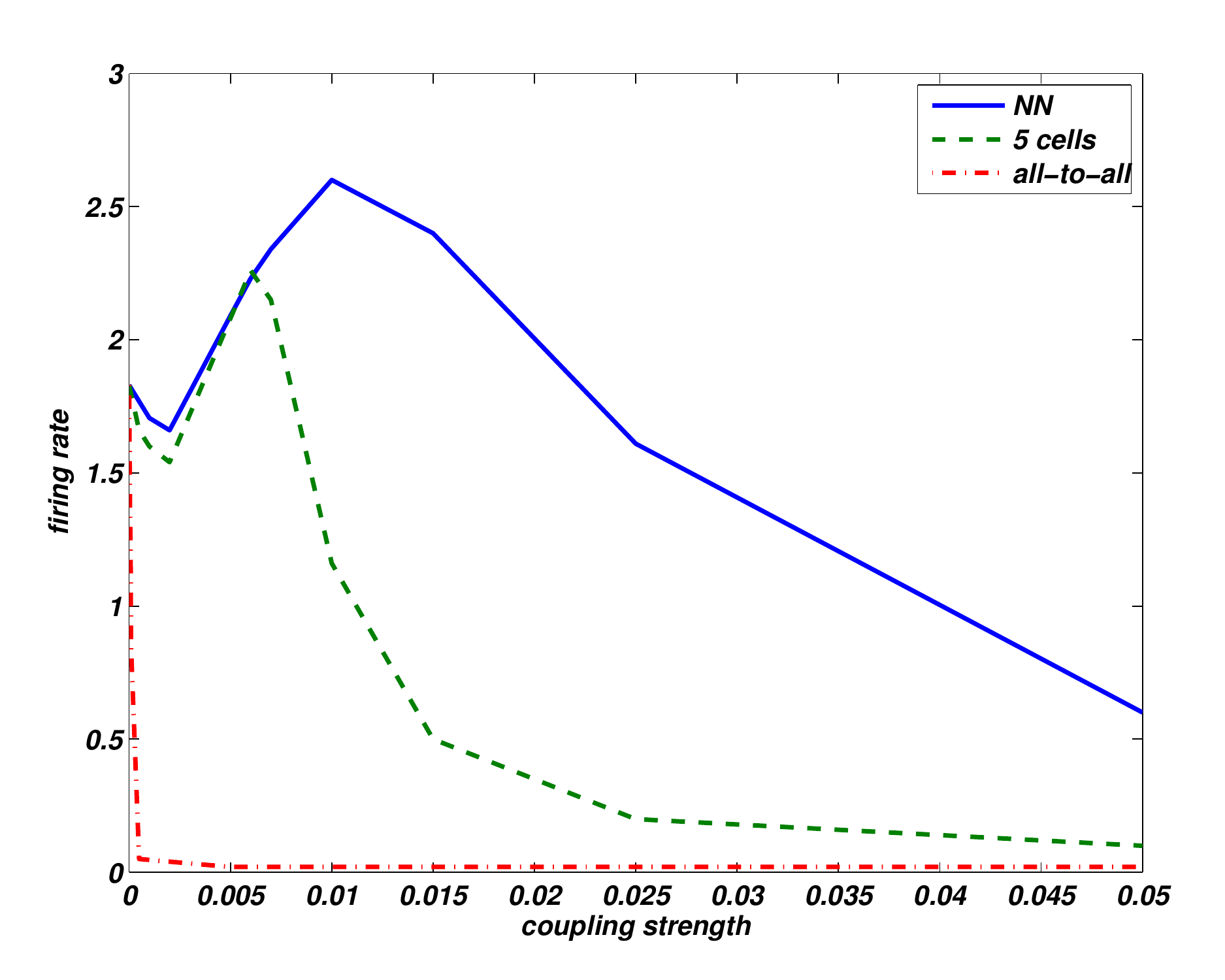}
{\bf b}\includegraphics[height=1.8in,width=2.0in]{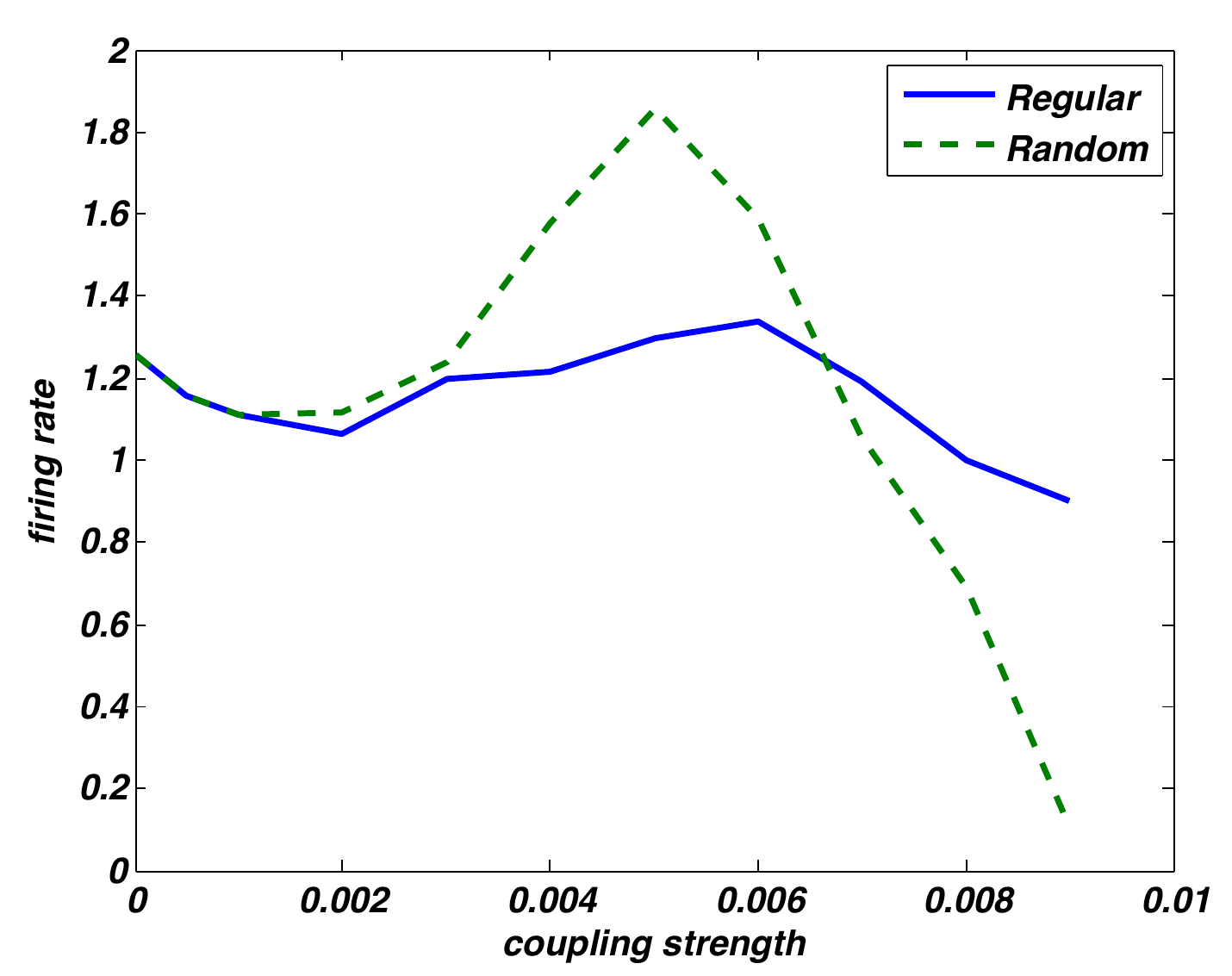}
{\bf c}\includegraphics[height=1.8in,width=2.0in]{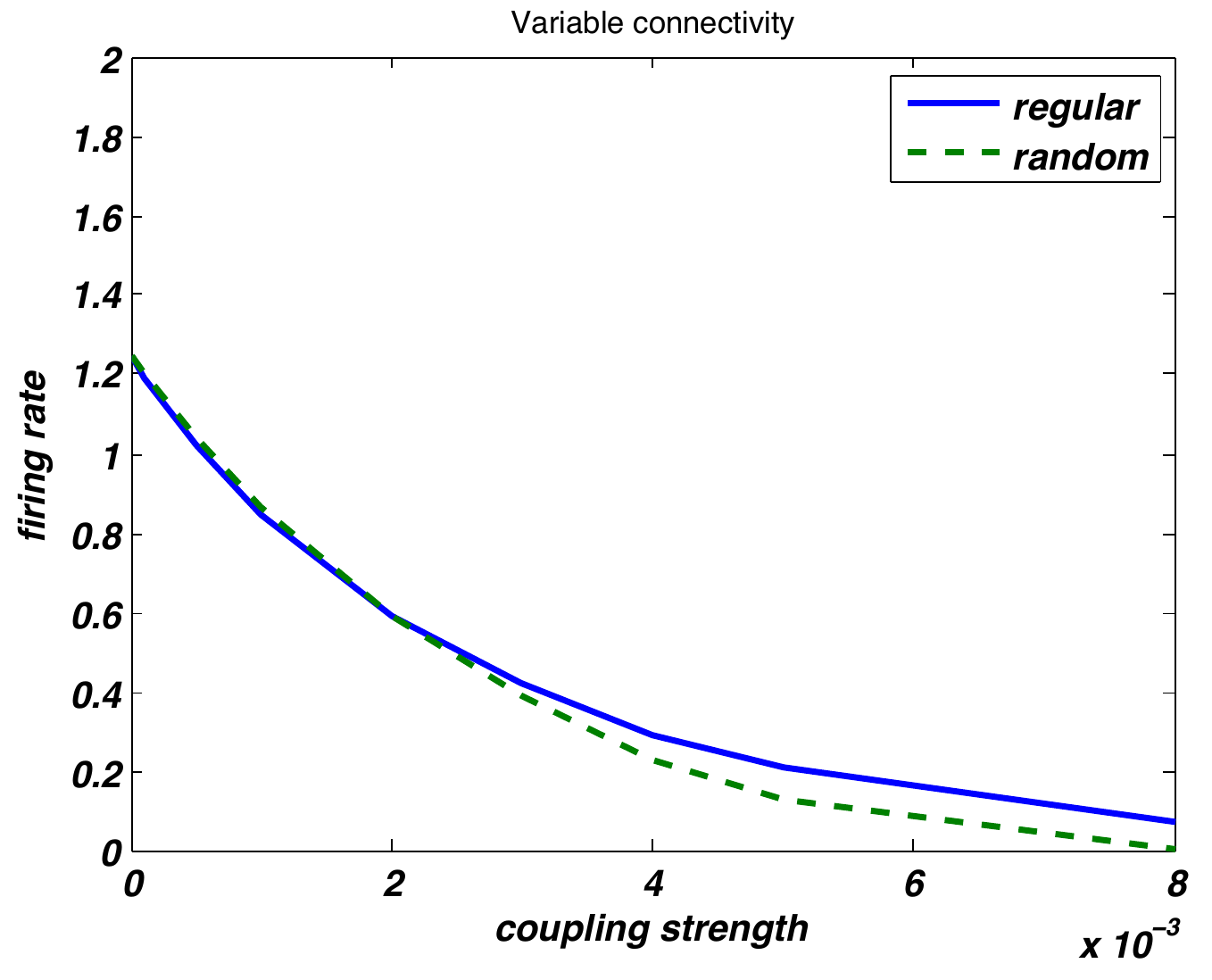}
\end{center}
\caption{
The fundamental relation of the rate of spontaneous activity 
and the coupling strength. The graphs in (a) are plotted for 
three coupling configurations: the nearest neighbor (dashed line),
the $2-$nearest neighbor coupling (solid line) and all-to-all coupling
(dash-dotted line) (see Examples \ref{ex.1}-\ref{ex.3}). The graphs in 
(b) are plotted for  the symmetric and random degree $4$ graphs
in solid and dashed lines respectively (see Example~\ref{ex.4}).
(c) The firing rate plot for the model, in which the coupling is turned 
off for values of the membrane potential above the firing threshold. The symmetric
(solid line) and random (dashed line) degree $4$ graphs are used for the two plots in (c).    
}\lbl{f.4}
\end{figure}

\section{Numerical experiments} \lbl{s3}
\setcounter{equation}{0}

The four parameters controlling the dynamics of the biophysical
model (\ref{1.4}) and (\ref{1.5}) are the excitability, 
the noise intensity, the coupling strength, and the network topology.
Assuming that the system is at a fixed distance from the bifurcation,
we study the dynamics of the coupled system  for sufficiently small
noise intensity $\sigma$. Therefore, the two remaining parameters
are the coupling strength and the network topology. 
We focus on the impact of the coupling strength on 
the spontaneous dynamics first. At the end of this section, 
we discuss the role of the network topology. The numerical experiments
of this section show that activity patterns generated by the network 
are effectively controlled by the variations of the coupling strength.
\begin{figure}
\begin{center}
{\bf a}\includegraphics[height=2.0in,width=2.5in]{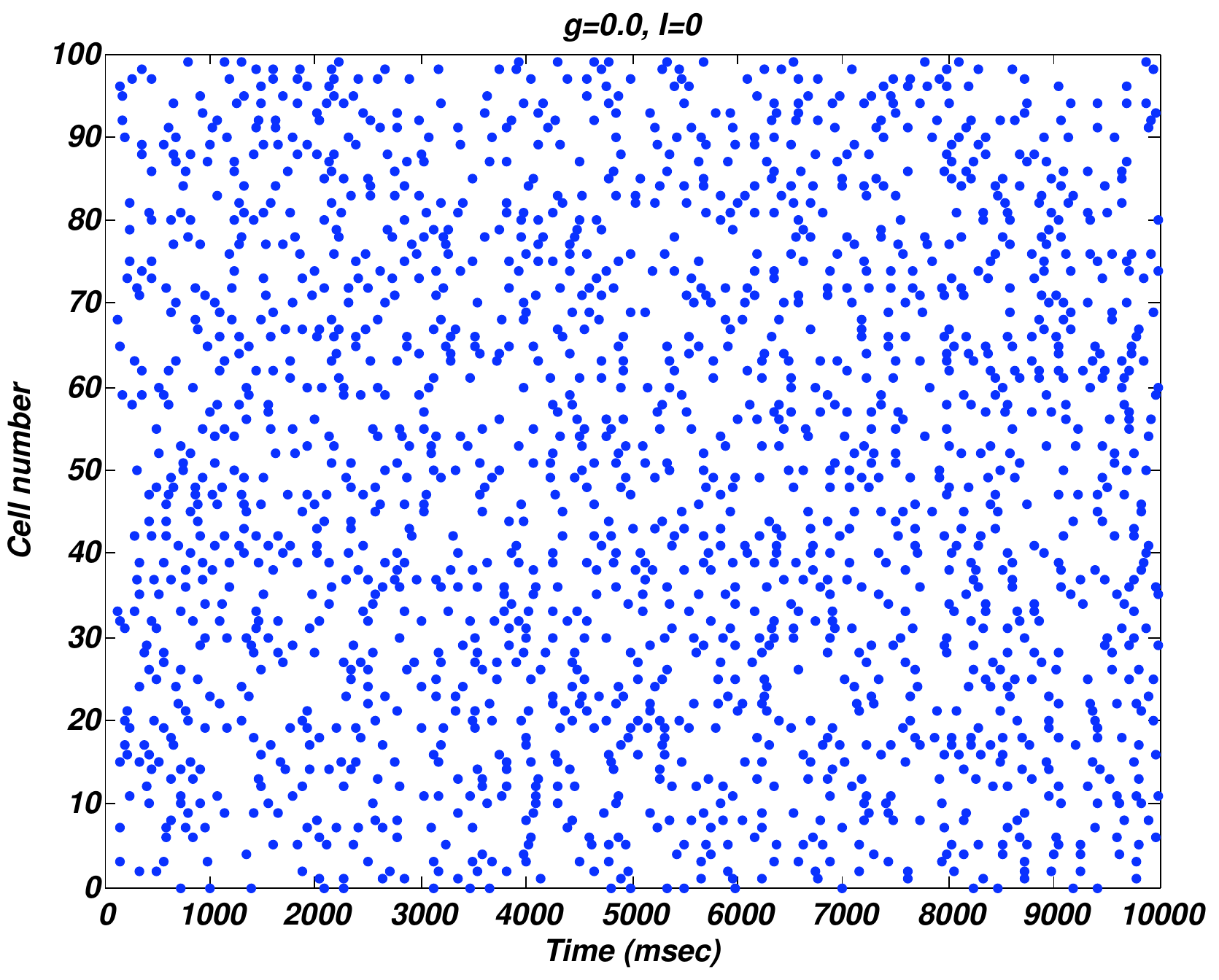}
{\bf b}\includegraphics[height=2.0in,width=2.5in]{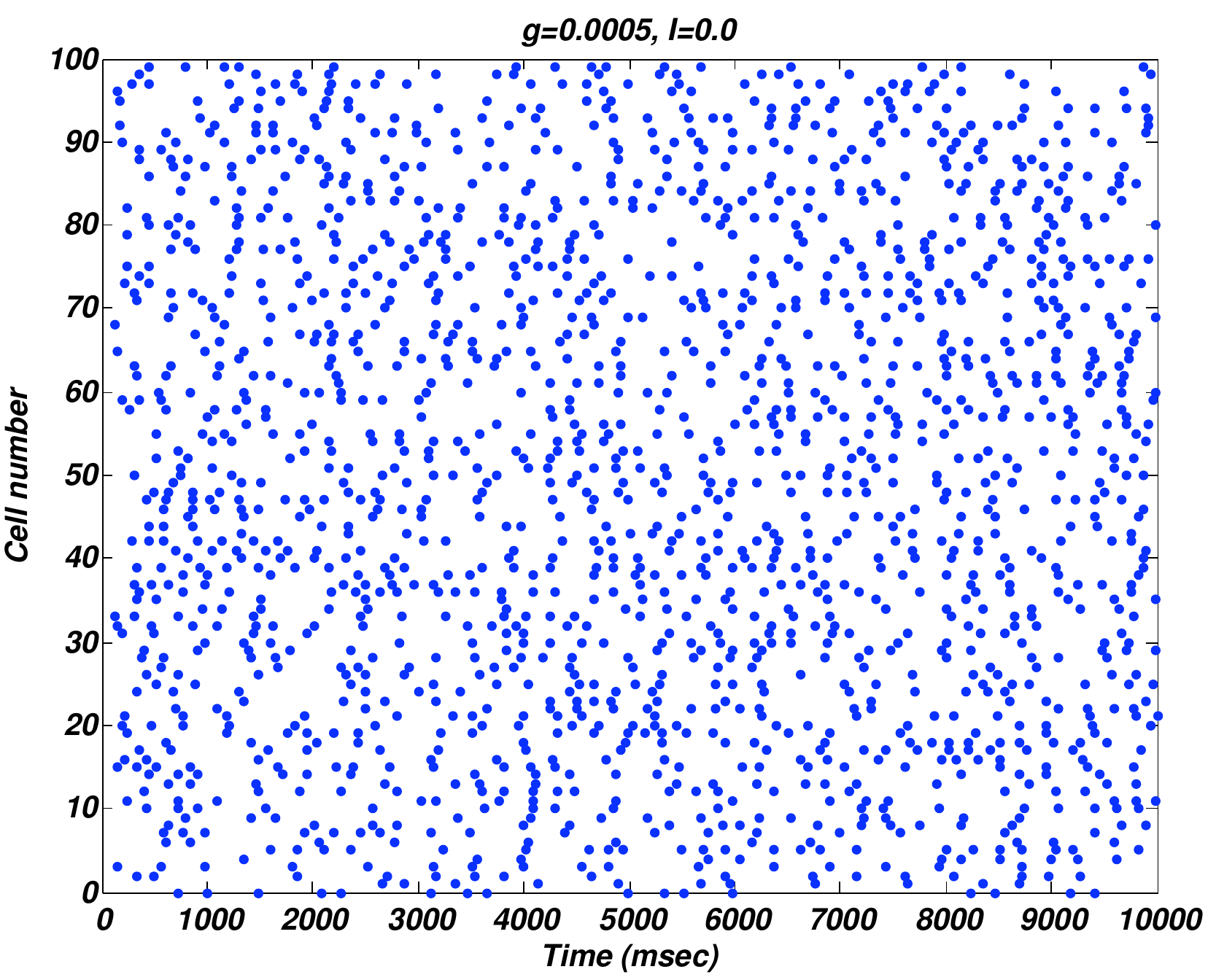} 
{\bf c}\includegraphics[height=2.0in,width=2.5in]{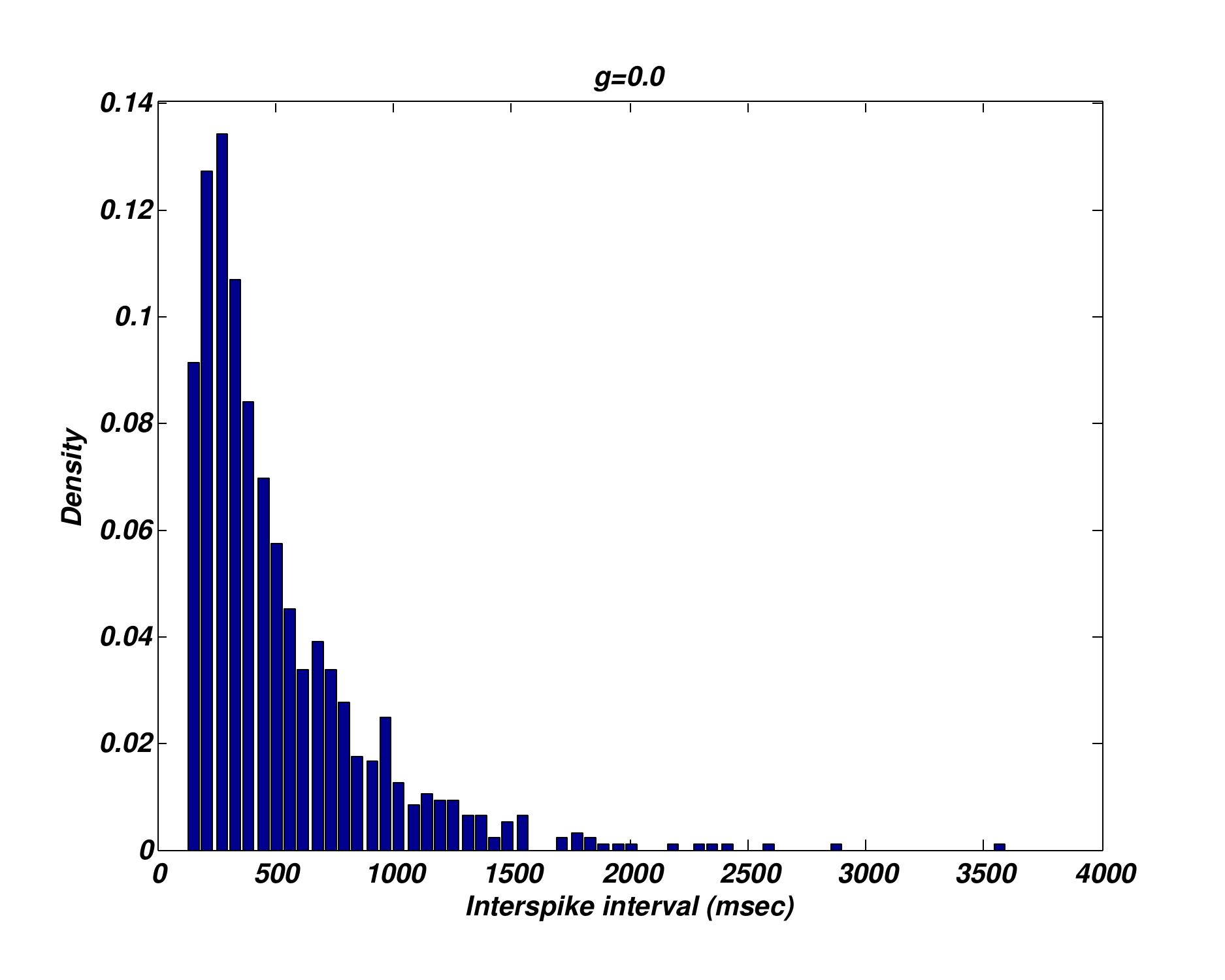}
{\bf d}\includegraphics[height=2.0in,width=2.5in]{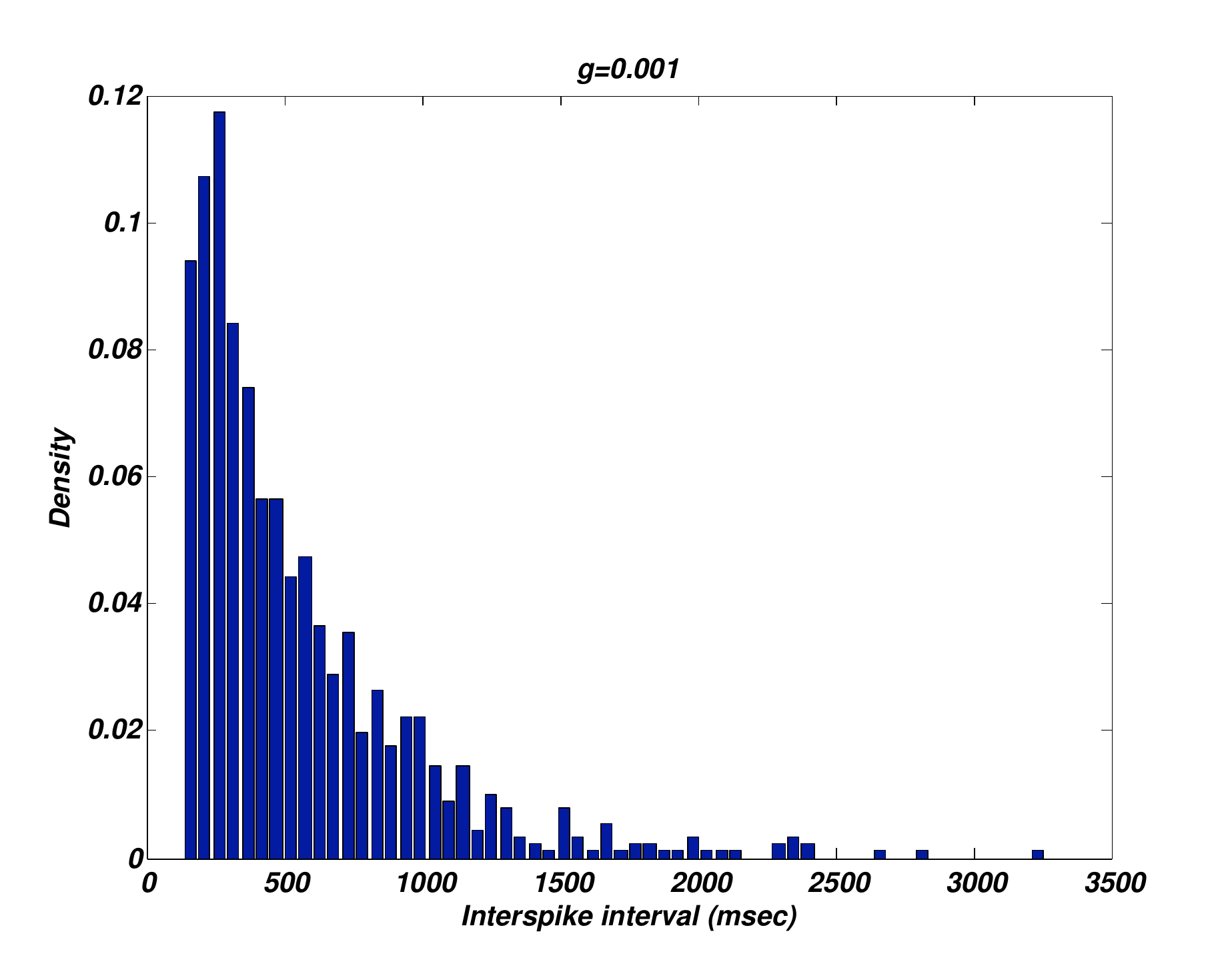}
\end{center}
\caption{Spontaneous activity in uncoupled (a) and weakly coupled (b)
networks. The corresponding distributions for the time intervals
between successive spikes are exponential (c,d) with a slightly 
heavier tail in the latter case.
}\lbl{f.5}
\end{figure}

\subsection{Three phases of spontaneous activity}
To measure the activity of the network for different values of the 
control parameters, we will use the average firing rate -
the number of spikes generated by the network per one neuron and per 
unit time. Fig.~\ref{f.4}a shows that the activity rate varies significantly
with the coupling strength. 
The three intervals of monotonicity of the activity rate plot reflect three 
main stages in the network 
dynamics en route
to complete synchrony: weakly correlated spontaneous spiking, formation of 
clusters and wave propagation, and synchronization.
We discuss these regimes in more detail below.
\begin{figure}
\begin{center}
{\bf a}\includegraphics[height=2.0in,width=2.5in]{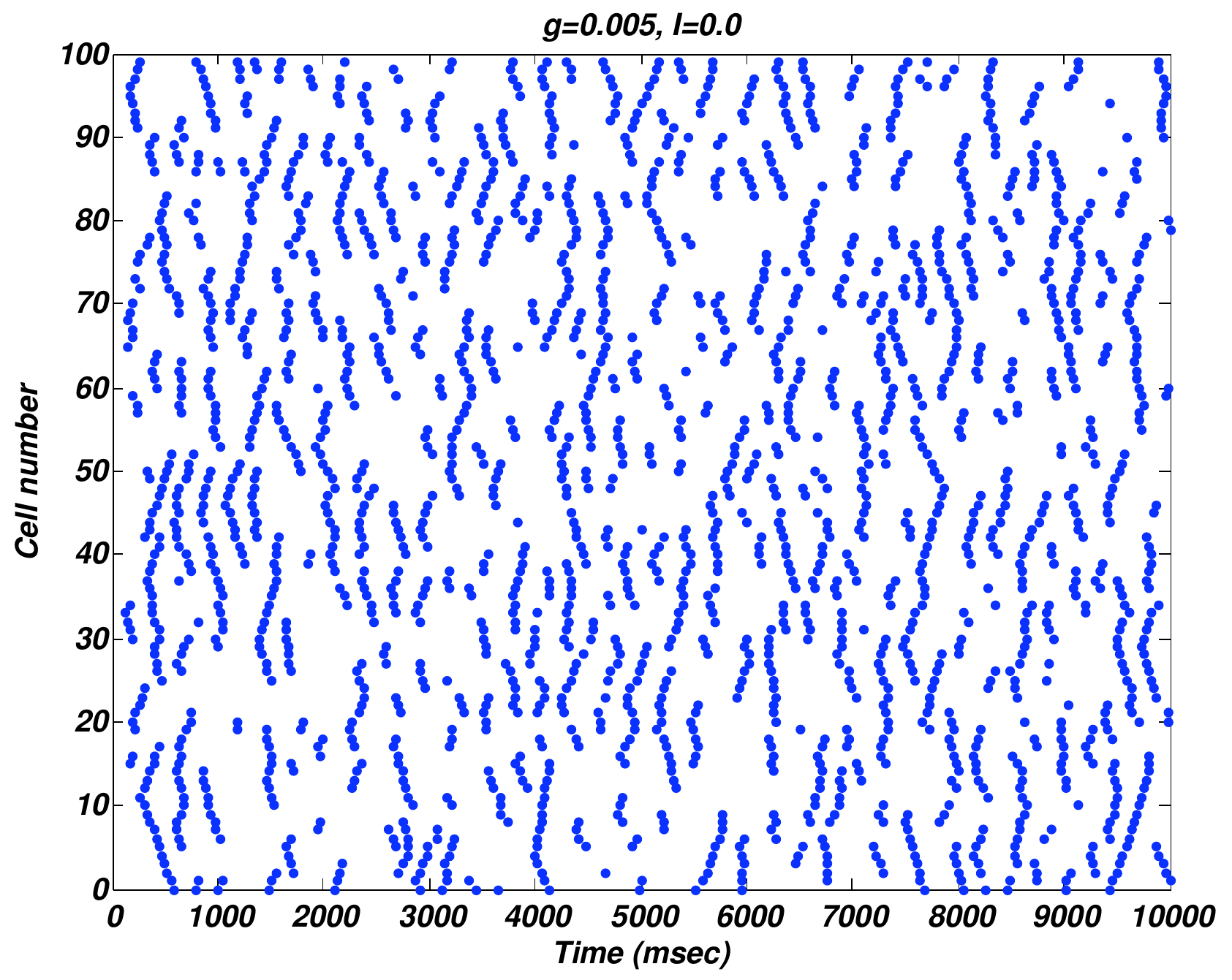}
{\bf b}\includegraphics[height=2.0in,width=2.5in]{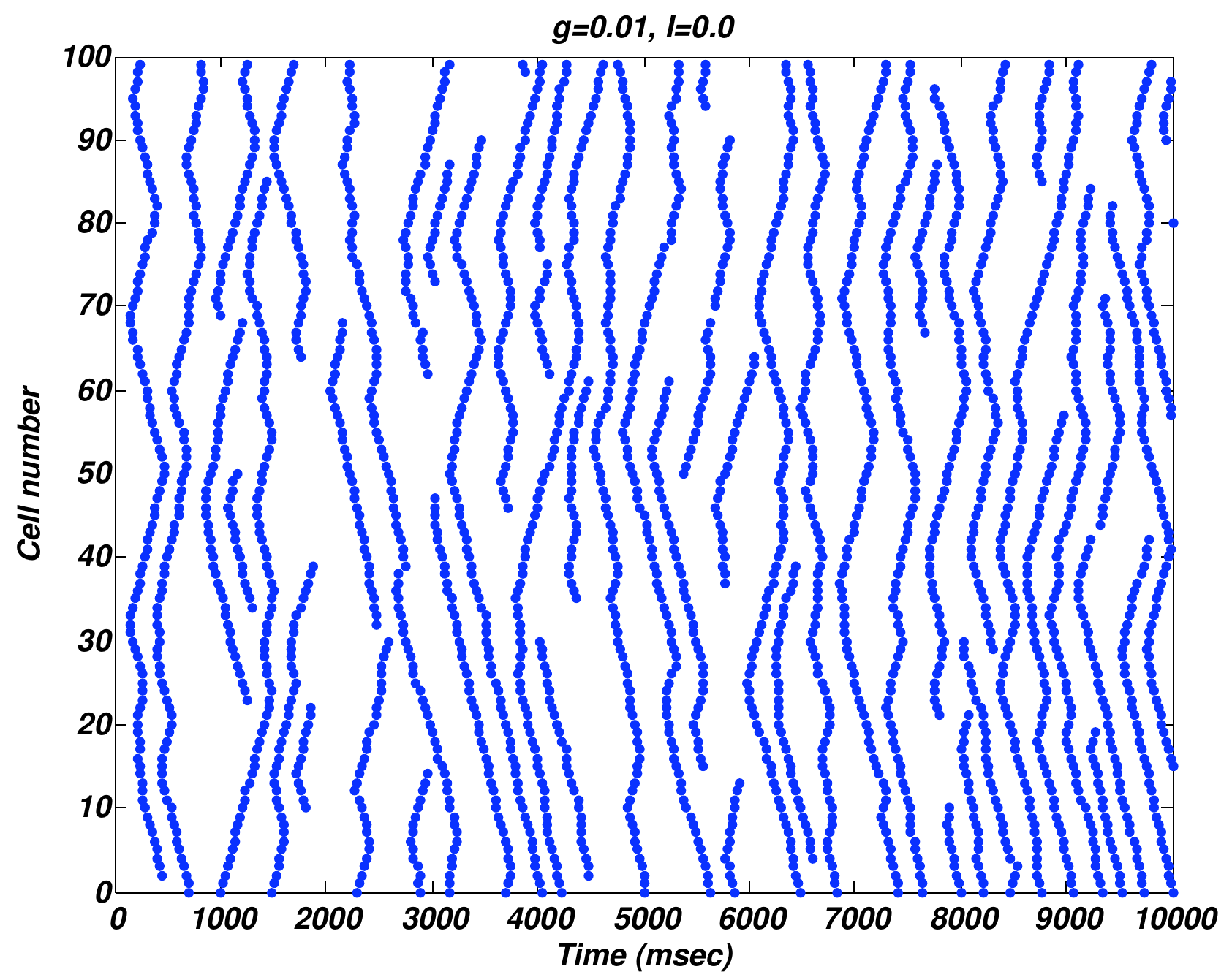}
{\bf c}\includegraphics[height=2.0in,width=2.5in]{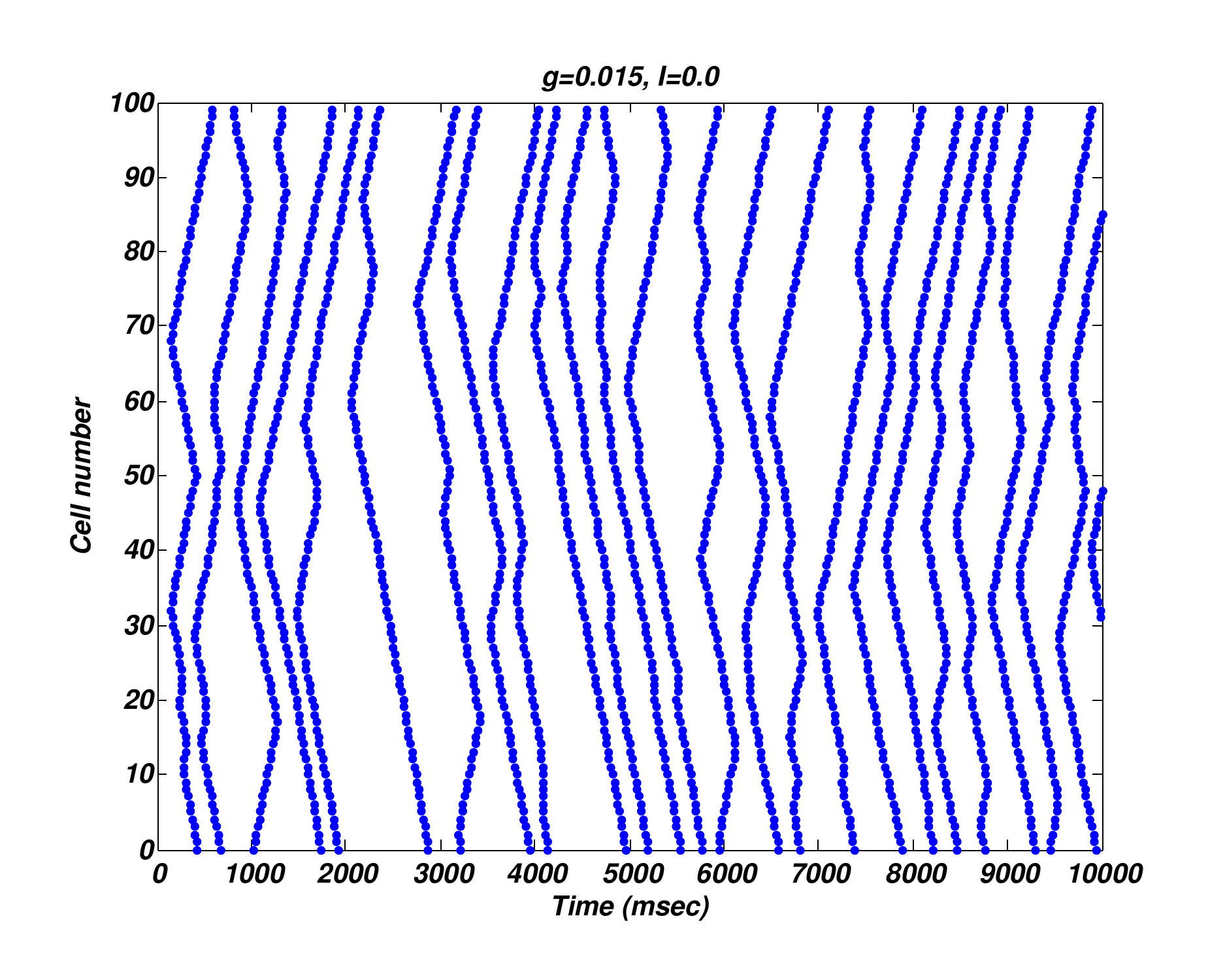}
{\bf d}\includegraphics[height=2.0in,width=2.5in]{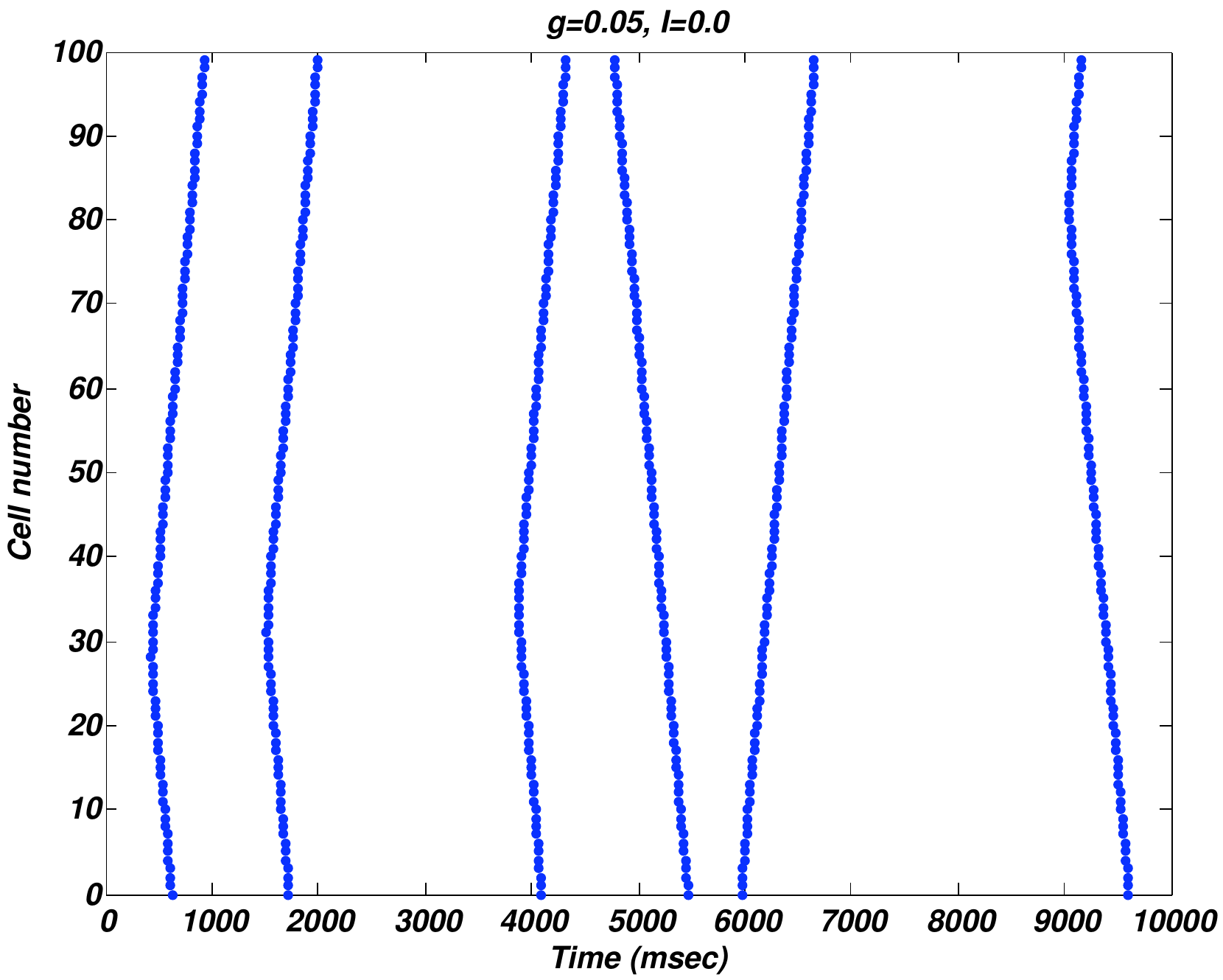}
\end{center}
\caption{Coherent structures in the weakly coupled network: a) clusters and short waves, 
b,c) robust waves, d) nearly
synchronous discharge. 
Networks shown  in Figures~\ref{f.5}-\ref{f.6a} are coupled through the nearest neighbor 
scheme.
}\lbl{f.6}
\end{figure}

\begin{figure}
\begin{center}
{\bf a}\includegraphics[height=2.0in,width=2.5in]{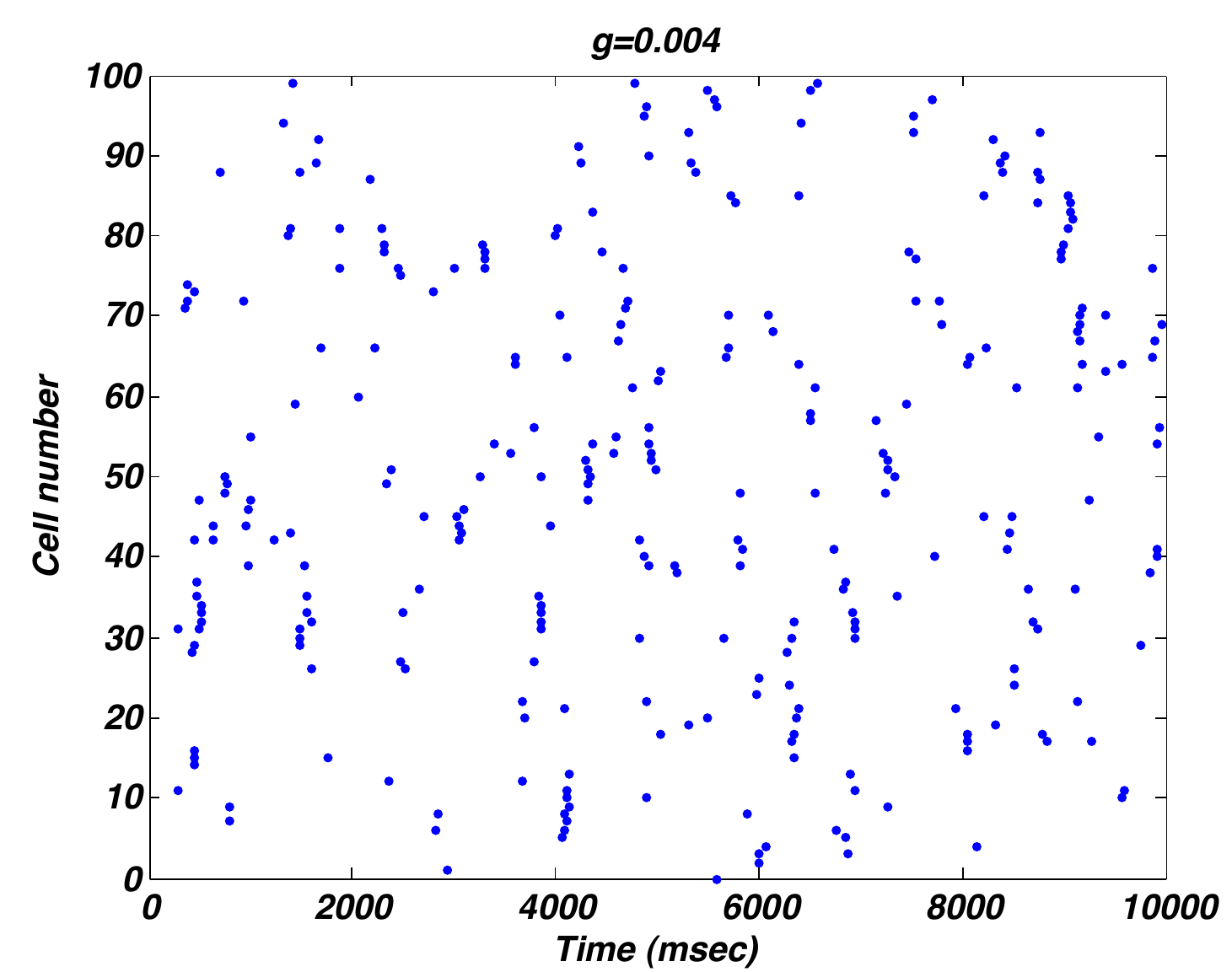}
{\bf b}\includegraphics[height=2.0in,width=2.5in]{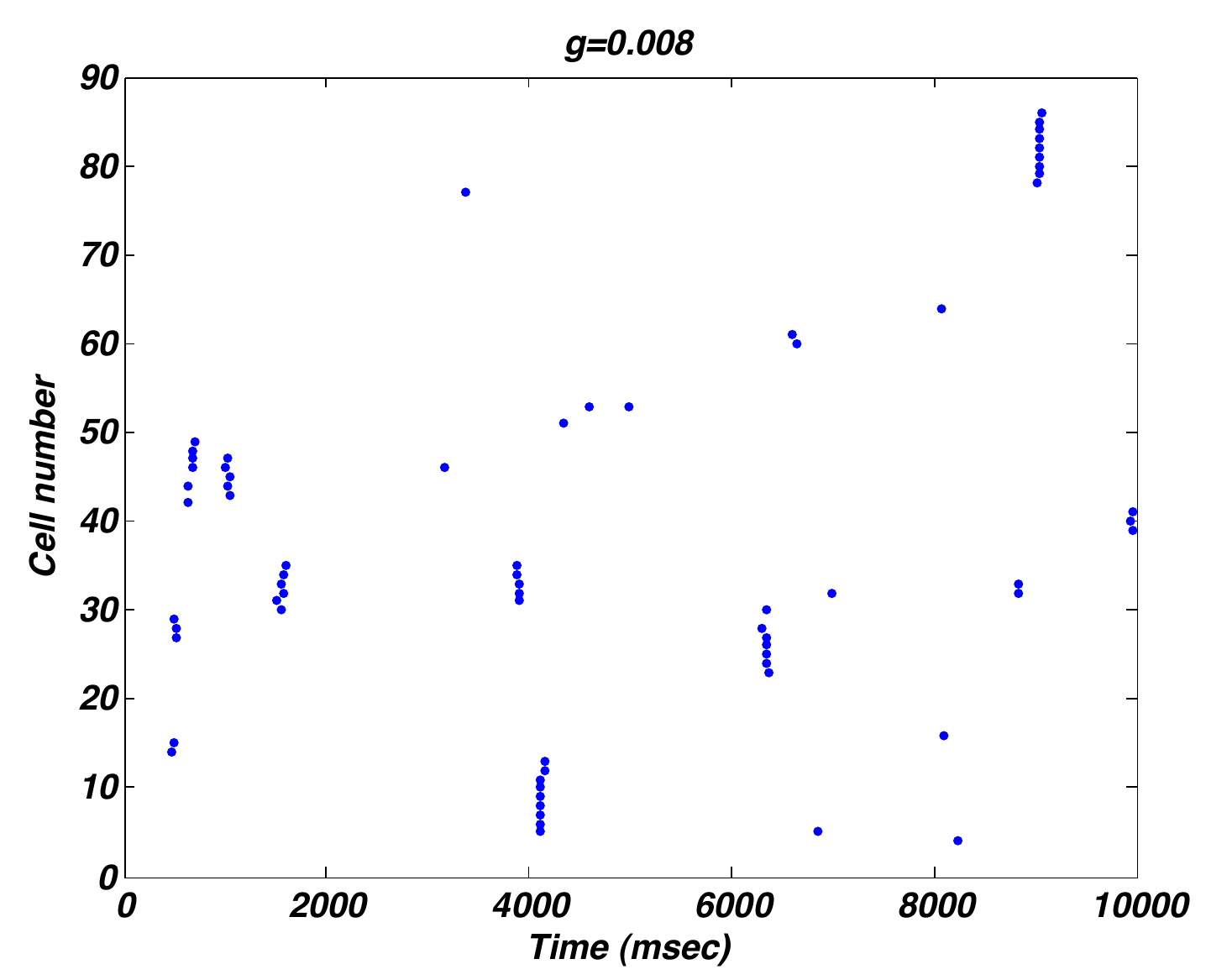}
\end{center}
\caption{Clusters generated by the modified model, in which the electrical current from a 
given cell is turned off once the cell has crossed the threshold (see text for details).
These experiments show that Factor A vs. B is responsible for forming clusters
in the weak coupling regime. 
}\lbl{f.6a}
\end{figure}

\noindent \textbf{Weakly correlated spontaneous spiking.} 
For $g>0$ sufficiently small, the activity retains the features
of spontaneous spiking in the uncoupled population. Fig.~\ref{f.5}b
shows no significant correlations between the activity of distinct cells 
in the weakly coupled network.
The distributions of the interspike intervals are exponential in
both cases (see Fig.~\ref{f.5} (c,d)).
There is an important change, however: the rate of firing goes down 
for increasing
values of $g\ge 0$ for small $g$. This is clearly seen from the graphs 
in Fig.~\ref{f.4}. The decreasing firing rate for very weak coupling
 can also be noted from the interspike interval distributions in 
Fig.~\ref{f.5}c,d: the density in Fig.~\ref{f.5}d has a heavier
tail.  
Thus, weak electrical coupling 
has a pronounced inhibitory (shunting) effect on the network dynamics: 
it drains the 
current from a neuron developing a depolarizing potential and redistributes
it among the cells connected to it. This  effect is stronger for networks
with greater number of connections. The three plots shown in Fig.~\ref{f.4}a
correspond to nearest-neighbor coupling, 
$2-$nearest neighbor coupling, and all-to-all coupling. Note that the slope 
at zero is steeper for networks with greater degree.

\noindent \textbf{Coherent structures.}
For increasing values of $g>0$ the system develops clusters, short waves, and
robust waves (see Fig.~\ref{f.6}).
The appearance of these spatio-temporal patterns starts in the middle of the 
first decreasing portion of the firing rate plot in Fig.~\ref{f.4}a and 
continues through the next (increasing) interval of monotonicity.
While patterns in Fig.~\ref{f.6} feature progressively increasing role 
of coherence in the 
system's dynamics, the dynamical mechanisms underlying cluster formation
and wave propagation are distinct. 
Factors A and B below identify two dynamical principles underlying pattern
formation in this regime.

\noindent
\textbf{Factor A:}\;{\it At the moment when one neuron fires
due to large deviations from the rest state, neurons
connected to it are more likely to be closer to the threshold
and, therefore, are more likely to fire within a short interval of time.}

\noindent
\textbf{Factor B:}\;{\it
When a neuron fires, it supplies neurons connected to it with depolarizing
current. If the coupling is sufficiently strong, the gap-junctional current 
triggers action potentials in these cells and the activity propagates through the network.
} 

Factor A follows from the variational interpretation of the spontaneous 
dynamics in weakly coupled networks, which we develop in Section~\ref{analysis}.
It is responsible for the formation of clusters and short waves, like
those shown in Fig.~\ref{f.6}a. To show numerically that that Factor A
(vs. Factor B) is responsible for the formation of clusters, 
we modified the model (\ref{1.4})
and (\ref{1.5}) in the following way. Once a neuron in the network has 
crossed the
threshold, we turn off the current that  it sends to the other neurons in the
network until it gets back close to the the stable fixed point.
We will refer to this model as the modified model
(\ref{1.4}) and (\ref{1.5}).  Numerical results for the modified model
in Fig.~\ref{f.6a}a,b,  show that clusters are formed as the result of the 
subthreshold dynamics, i.e., are due to Factor A. Factor B becomes dominant 
for stronger coupling. It results in robust waves with constant speed of 
propagation. The mechanism of the wave propagation is essentially deterministic
and is well known from the studies of waves in 
excitable systems (cf. \cite{Kee87}). 
However, in the presence of noise, the excitation and termination of waves 
become random (see Fig.~\ref{f.6}(b,c)).

\noindent\textbf{Synchrony.}
The third interval of monotonicity in the graph of the firing rate vs. the
coupling strength is decreasing (see Fig.~\ref{f.4}a). 
It features synchronization, the final dynamical
state of the network. In this regime, once one cell
crosses the firing threshold the entire network fires in unison. 
The distinctive feature of this
regime is a rapid decrease of the firing rate for increasing $g$ 
(see Fig.~\ref{f.4}a). The slowdown of firing in the strong coupling
regime was studied in \cite{medvedev09} 
(see also  \cite{medvedev10, medvedev10b, MZ11}).
When the coupling is strong the effect of noise on the network dynamics
is diminished by the dissipativity of the coupling operator.
The reduced effect of noise results 
in the decrease of the firing rate.
In \S\ref{topology}, we present analytical estimates 
characterizing denoising by electrical coupling for the present model.
\begin{figure}
\begin{center}
\textbf{a}\includegraphics[height=2.2in,width=2.7in]{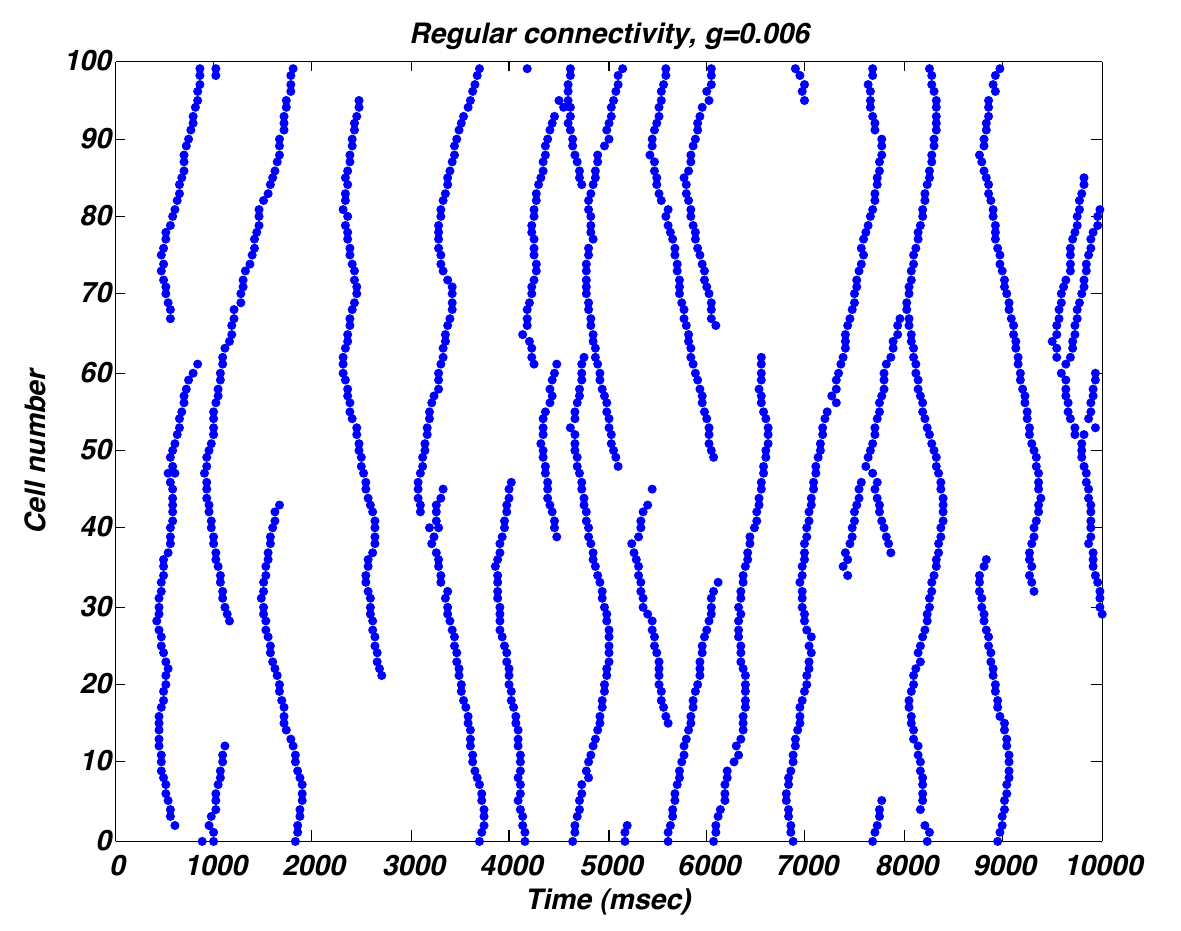}
\textbf{b}\includegraphics[height=2.2in,width=2.7in]{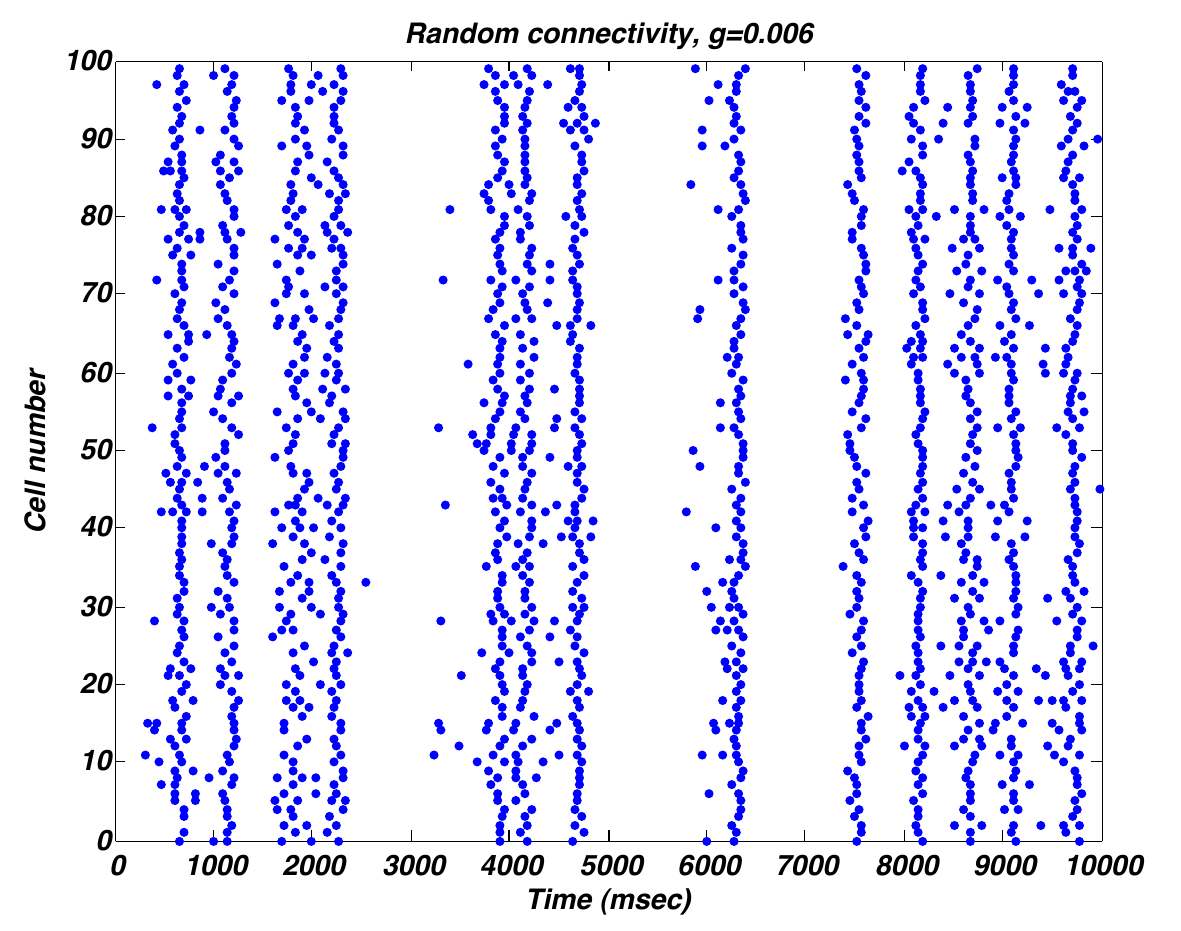}
\end{center}
\caption{Spontaneous activity patterns generated by regularly 
(a) and randomly (b) connected 
degree-4 networks (cf. Example~\ref{ex.4}) for the same value of 
the coupling strength $g=0.006$. The randomly connected network is 
already synchronized (b), while the regular network is en route to 
synchrony (a).  
}\lbl{f.8}
\end{figure}

\subsection{The role of the network topology}
All connected networks of excitable elements (regardless of the connectivity pattern) 
undergo the three dynamical regimes, 
which we identified above for weak, intermediate, and strong coupling. The topology 
becomes important for quantitative description of the activity patterns. 
In particular, the topology affects the boundaries between different phases.
We first discuss the role of topology for the onset of synchronization.
The transition to synchrony corresponds to the beginning of the third phase and 
can be approximately identified with the location of the point of maximum on the 
firing rate plot (see Fig.~\ref{f.4}a,b). The comparison of the plots for $1-$ and 
$2-$nearest-neighbor coupling schemes shows that the onset of synchrony takes 
place at a smaller value of $g$ for the latter network. This illustrates a general 
trend: networks with greater number of connections tend to have better
synchronization properties. However, the degree is not the only 
structural property of the graph that affects synchronization. The connectivity 
pattern is important as well. Fig.~\ref{f.8} shows that a randomly connected 
degree $4$ network synchronizes faster than its symmetric counterpart
(cf. Example~\ref{ex.4}). 
The analysis in \S\ref{strong} shows that the point of transition to synchrony 
can be estimated using the algebraic connectivity of the graph  
$\mathfrak{a}$. Specifically, the network is synchronized,
if
$
\gamma > \mathfrak{a}^{-1},
$
where $\gamma$ stands for the coupling strength in the rescaled nondimensional 
model. The algebraic connectivity is easy to compute numerically. 
For many graphs with symmetries
including those in Examples~\ref{ex.1}-\ref{ex.3}, the algebraic connectivity
is known analytically. On the other hand, there are effective asymptotic estimates
of the algebraic connectivity available for certain classes of graphs that are
important in applications, such as random graphs \cite{Fri08} and 
expanders \cite{Hoory06}.
The algebraic connectivities  of the graphs in Examples~\ref{ex.1}-\ref{ex.2}
$\mathfrak{a}=O(n^{-2})$ tend to zero as $n\to\infty$. Therefore, for such networks
one needs to increase the strength of coupling significantly to maintain synchrony 
in networks growing in size. This situation is typical for symmetric or almost symmetric
graphs.
In contrast, it is known that for the random graph from Example~\ref{ex.4}
the algebraic connectivity is bounded away from zero (with high probability)
as $n\to\infty$ \cite{Fri08,Hoory06}. Therefore, one can guarantee synchronization 
in dynamical networks on such graphs using  finite coupling strength 
when the size of the network grows without bound. This counter-intuitive 
property is intrinsic to networks on expanders, sparse well connected graphs \cite{Hoory06, Sar04}. 
For a more detailed discussion of the role of network topology in synchronization,
we refer the interested reader to Section~$5$ in \cite{medvedev10b}.

The discussion in the previous paragraph suggests that connectivity is important
in the strong coupling regime. It is interesting that to a large extent the 
dynamics in the weak coupling regime remains unaffected by the connectivity. 
For instance, the firing rate plots for the random and symmetric
degree-$4$ networks (Example~\ref{f.4}) shown in Fig.~\ref{f.4}b coincide
over an interval in $g$ near $0$. Furthermore, the plots for the same pair
of networks based on the modified model (\ref{1.4}) and (\ref{1.5}) are almost
identical, regardless the disparate connectivity patterns underlying these
networks.  
The variational analysis in \S\ref{weak}
shows that, in the weak coupling regime, to leading order the firing rate 
of the network depends only on the number of connections between cells.
The role of the connectivity in shaping  network dynamics increases in
the strong coupling regime.

\section{The variational analysis of spontaneous dynamics}\lbl{analysis}
\setcounter{equation}{0}
In this section, we analyze dynamical regimes of the coupled system (\ref{1.4}) 
and (\ref{1.5}) under the variation of the coupling strength. In \S\ref{center},
we derive an approximate model using the center manifold reduction. In \S\ref{exit-problem},
we relate the activity patterns of the coupled system to the minima of a certain
continuous function on the surface of an $n-$cube. The analysis of the  minimization problem
for weak, strong, and intermediate coupling is used to characterize the dynamics of the 
coupled system in these regimes.

\subsection{The center manifold reduction}\lbl{center}
In preparation for the analysis of the coupled system (\ref{1.4}) and (\ref{1.5}),
we approximate it by a simpler system using the center manifold
reduction \cite{CH82, GH}. To this end, we first review the bifurcation structure 
of the model.
Denote the equations governing the deterministic dynamics of a single 
neuron by 
\be\lbl{local}
\dot x=\mathsf{f}(x,\mu),
\ee
where $x\in\R^d$ and $\mathsf{f}:\R^d\times\R^1\to\R^d$ is a smooth function
and $\mu$ is a small parameter, which controls the distance of (\ref{local}) from the
saddle-node bifurcation. 
\begin{assume}\lbl{SN}
Suppose that at $\mu=0$, 
the unperturbed problem (\ref{local}) 
has a nonhyperbolic equilibrium at the origin such that $D\mathsf{f}(0,0)$ has a single
zero eigenvalue and the rest of the spectrum lies to the left of the imaginary
axis. Suppose further that at $\mu=0$ there is a homoclinic orbit to $O$ 
entering the origin along the $1D$ center manifold.
\end{assume}
\begin{figure}
\begin{center}
\includegraphics[height=2.0in,width=2.5in]{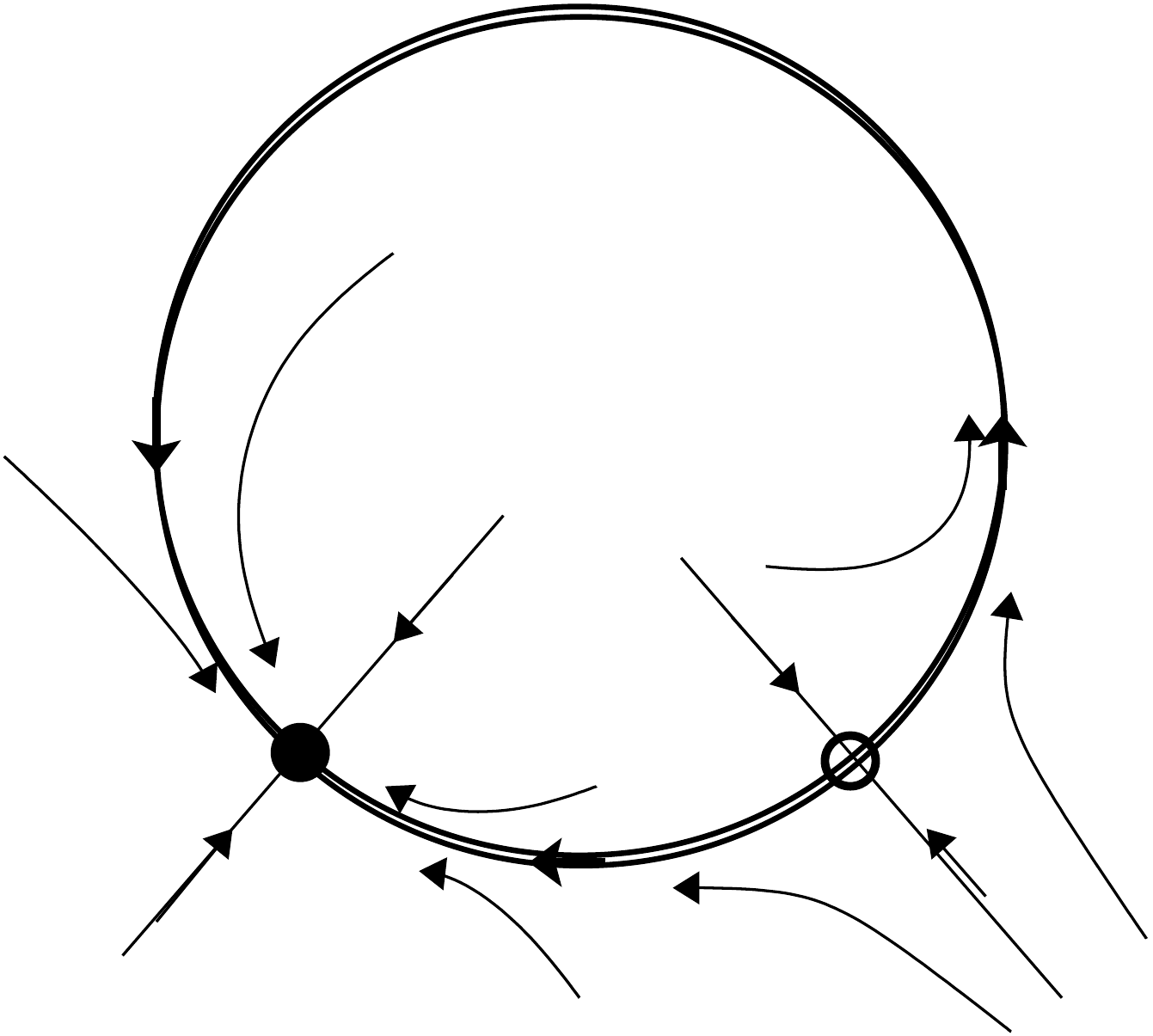}
\end{center}
\caption{ Local system (\ref{local}) is near the saddle-node
on an invariant circle bifurcation.
}
\lbl{excitable}
\end{figure}
Then under appropriate nondegeneracy and transversality conditions on the local
saddle-node bifurcation at $\mu=0$, for $\mu$ near zero
the homoclinic orbit is transformed into either a unique asymptotically stable
periodic orbit or to a closed invariant curve $C_\mu$ having two equilibria:
a node and a saddle \cite{CH82} (Fig.~\ref{excitable}). Without loss of generality, 
we assume that the latter  case is realized for small positive $\mu$, and the 
periodic orbit exists for negative $\mu$. 
Let $\mu>0$ be a sufficiently small fixed number, i.e., (\ref{local})
is in the excitable regime (Fig. \ref{excitable}). 
For simplicity, we assume that the stable node near the origin is the only attractor 
of (\ref{local}).

We are now in a position to formulate our assumptions on the coupled system. 
Consider $n$ local systems (\ref{local}) that are placed at the nodes of the 
connected graph $G=(V(G),E(G)),\; |V(G)|=n,$ and coupled electrically:
\be\lbl{coup}
\dot X = \mathsf{F}(X,\mu)-g(L\otimes J)X +\sigma (I_n\otimes P)\dot W,
\ee
where $X=(x^{(1)},x^{(2)},\dots, x^{(n)})^\t\in\R^d\times\dots\times\R^d=\R^{nd},$
$\mathsf{F}(X,\mu)=\left(\mathsf{f}(x^{(1)},\mu),\mathsf{f}(x^{(2)},\mu),\dots, 
\mathsf{f}(x^{(n)},\mu)\right)^\t$,
$I_n$ is an $n\times n$ identity matrix, $P\in\R^{d\times d}$, and 
$L\in\R^{n\times n}$ is the Laplacian 
of $G$. Matrix $J\in\R^{d\times d}$ defines the linear combination of the local
variables engaged in coupling. In the the neuronal network model above, 
$J=\mathrm{diag}(1,0)$. 
Parameters $g$
and $\sigma$ control the coupling strength and the noise intensity respectively.
$\dot W$ is a Gaussian white noise process in $\R^{nd}$.
The local systems are taken to be identical for simplicity.
The analysis can be extended to cover nonhomogeneous networks.

We next turn to the center manifold reduction of (\ref{coup}).
Consider (\ref{coup}$)_0$ (the zero subscript refers to $\sigma=0$)
for $\mu=g=0$. By our assumptions on the 
local system (\ref{local}), $D\mathsf{f}(0,0)$ has a $1D$ kernel. Denote
\be\lbl{kerA}
e\in\ker D\mathsf{f}(0,0)/\{0\} \;\mbox{and}\; p\in\ker (D\mathsf{f}(0,0))^\t\;\mbox{such that}\; 
p^\t e=1.
\ee
By the center manifold theorem,
there is a neighborhood of the origin in the phase space of (\ref{coup}),
$B$, and $\delta>0$ such that 
for $|\mu|<\delta$ and $|g|<\delta$, in $B$, there exists an attracting 
locally invariant $n-$dimensional slow manifold 
$\mathcal{M}_{\mu,g}$. The trajectories 
that remain in $B$ for sufficiently long time can be approximated by those lying in 
$\mathcal{M}_{\mu,g}$. Thus, the dynamics of (\ref{coup}$)_0$ 
can be reduced to $\mathcal{M}_{\mu,g}$,
whose dimension is $d$ times smaller than that of the phase space of (\ref{coup}$)_0$.
The center manifold reduction is standard. Its justification relies on the 
Lyapunov-Schmidt method and Taylor expansions (cf. \cite{CH82}). Formally,
the reduced system is obtained by projecting (\ref{coup}$)_0$ onto the
center subspace of (\ref{coup}$)_0$ for $\mu=g=0$
(see \cite{Kuz98}):
\be\lbl{red}
\dot y=a_1 y^2 -a_2\mu - a_3g Ly+ O(|y|^3,\mu^2,g^2),
\ee
where $y=(y_1,y_2,\dots,y_n)\in\R^n,$
$y^2:=(y^2_1,y^2_2,\dots,y^2_n);$ provided that
the following nondegeneracy conditions hold 
\begin{eqnarray}
\lbl{a1}
a_1 &=& {1\over 2} {\partial^2\over \partial u^2} p^\t \mathsf{f}(ue,0)\left|_{u=0}\right.\neq 0,\\
\lbl{a2}
a_2 &=&-{\partial \over \partial\mu} p^\t \mathsf{f}(0,\mu)\left|_{\mu=0}\right.\neq 0,\\ 
\lbl{a3}
a_3 &=& p^\t Je\neq 0.
\end{eqnarray}
Conditions (\ref{a1}) and (\ref{a2}) are the nondegeneracy and transversality
conditions of the saddle-node bifurcation in the local system 
(\ref{local}). 
Condition (\ref{a3}) guarantees that the projection of the coupling 
onto the center subspace is not trivial. 
All conditions are open. Without loss of generality, assume that nonzero
coefficients $a_{1,2,3}$ are positive.

Next, we include the random perturbation in the reduced model.
Note that near the saddle-node bifurcation ($0<\mu\ll 1$),
the vector field of (\ref{coup}$)_0$  is much stronger 
in the directions transverse to $\mathcal{M}_{\mu,g}$ than in the tangential directions.
The results of the geometric theory of randomly perturbed fast-slow
systems imply that the trajectories
of (\ref{coup}) with small positive $\sigma$ that start close 
to the node of (\ref{coup}$)_0$ remain in a small neighborhood
of $\mathcal{M}_{\mu,g}$ on finite intervals of time  with overwhelming 
probability (see \cite{BG} for specific estimates).
To obtain the leading order approximation of the stochastic system 
(\ref{coup}) near the slow manifold, we project the random perturbation
onto the center subspace of (\ref{coup}$)_0$ for $\mu=g=0$
and add the resultant term to the reduced equation (\ref{red}):
\be\lbl{rednoise}
\dot y=a_1 y^2-a_2\mu - a_3 g Ly+\sigma B\dot W+\dots,\; B=I_n\otimes (p^\t P)\in \R^{n\times nd}.
\ee
We replace $B\dot W$ by identically distributed $a_4 \dot w$, where $\dot w$ is
a white noise process in $\R^n$ and $a_4=|P^\t p|$. Here, $|\cdot|$ stands for the Euclidean
norm of $P^\t p\in \R^d$.
After rescaling the resultant equation and ignoring the higher order terms, 
we arrive at the following reduced model
\be\lbl{rescale}
\dot z=z^2-\mathbf{1_n} - \gamma Lz+ \sigma\dot w,
\ee
where $w$ stands for a standard Brownian motion in $\R^n$
and $\mathbf{1_n}=(1,1,\dots,1)\in\R^n.$
Here, with a slight abuse of notation, we continue to use
$\sigma$ to denote the small parameter in the rescaled system.
In the remainder of this paper,
we analyze the reduced model (\ref{rescale}). 

\subsection{The exit problem}\lbl{exit-problem}
In this subsection, the problem of identifying most likely
dynamical patterns generated by (\ref{coup}) is
reduced to a minimization problem for a smooth
function on the surface of the unit cube.
  
Consider the initial value problem for (\ref{rescale})
\be\lbl{diffusion}
\dot z=\mathbf{f}(z) - \gamma Lz+\sigma \dot w, \; L=H^\mathsf{T}H,\; 
z(0)=z_0\in D\subset\R^n,
\ee
where 
\be\lbl{definef}
\mathbf{f}(z)=(f(z_1), f(z_2),\dots, f(z_n)),\;\;f(\xi)=\xi^2-1,
\ee
and
\be\lbl{define-D}
D=\{ z=(z_1, z_2,\dots,z_n): -2-b < z_i < 1, i\in [n]:=\{1,2,\dots,n\}\},
\ee
where auxiliary parameter $b>0$ will be specified later.
Let 
\be\lbl{boundary}
\partial^+ D =\left\{ z=(z_1,z_2,\dots, z_n):\; z\in \bar D\;
\&\;(\exists i\in [n]\; z_i=1) \right\},
\ee
denote a subset of the boundary of D, $\partial D$. If $z(\tau)\in\partial^+ D$, then at least one 
of the neurons in the network is at the firing threshold. It will be shown below that 
the trajectories of  (\ref{diffusion}) exit from $D$ through $\partial^+ D$
with probability $1$ as $\sigma\to 0$,  provided $b>0$ is sufficiently large.\footnote[1]{Positive 
parameter $b$ in the definition of $D$ (cf.~(\ref{define-D})) is used 
to exclude the possibility of exit from $D$ through $\partial D\setminus \partial^+D$.}
Therefore, the statistics of the first exit time
\be\lbl{tau}
\tau=\inf~\{t>0:~z(t)\in\partial D \}
\ee
and the distribution of the location of the points of exit $z(\tau)\in\partial^+ D$
characterize the statistics of the interspike intervals and the most probable firing
patterns of (\ref{1.1}) and (\ref{1.2}), respectively.
The Freidlin-Wentzell theory of large deviations \cite{FW} yields  
the asymptotics of $\tau$ and $z(\tau)$ for small $\sigma>0$.

To apply the large deviation estimates  to the problem at hand, we rewrite
(\ref{diffusion}) as a randomly perturbed gradient system
\be\lbl{gradient}
\dot z=-{\partial\; \over\partial z} U_\gamma (z)+\sigma\dot w_t,
\ee
where
\begin{equation}\lbl{func-1}
U_\gamma(z) = {\gamma\over 2} \langle Hz,Hz\rangle +\Phi(z), \; 
\Phi(z) =\sum_{i=1}^n F(z_i), \; F(\xi)={2\over 3}+\xi-{1\over 3}\xi^3,
\end{equation}
where $\langle\cdot,\cdot\rangle$ stands for the inner product in $\R^{n-1}$.
The additive constant $2/3$ in the definition of the potential function 
$F(\xi)$ is used to normalize the value of the potential at the local
minimum $F(-1)=0$.

The following theorem summarizes the implications of the large deviation
theory for (\ref{gradient}). 
\begin{thm}\lbl{main}
Let $\bar z^{(1)},\bar z^{(2)},\dots, \bar z^{(k)}, \;k\in\mathbb{N},$
denote the points of minima of $U_\gamma(z)$ on $\partial D$ 
$$
U_\gamma(\bar z^{(i)})=\bar U:=\min_{z\in\partial D}U_\gamma(z),\;i=1,2,\dots,k,
$$
and $\bar Z=\bigcup_{i=1}^k\{\bar z^{(i)}\}$. Then for any $z_0\in D$ and $\delta>0$, 
\begin{eqnarray}\lbl{F1}
\mathbf{A)} & \lim_{\sigma\to 0} \P_{z_0}\{ \rho(z(\tau),\bar Z)<\delta\}=1,\\
\lbl{F2}
\mathbf{B)} &\lim_{\sigma\to 0} \sigma^2 \ln \E_{z_0} \tau= \bar U,\\
\lbl{F3}
\mathbf{C)} &\lim_{\sigma\to 0} \P_{z_0} \left\{ \exp\{\sigma^{-2}(\bar U-h)\}
<\tau< \exp\{\sigma^{-2}(\bar U+h)\}\right\}=1,\;\forall h>0,
\end{eqnarray}
where $\rho(\cdot, \cdot)$ stands for the distance in $\R^n$.
\end{thm}
The statements A)-C) can be shown by adopting the proofs of Theorems 2.1, 
3.1, and 4.1 of Chapter~4 of \cite{FW} to the case of the action functional
with multiple minima. 

Theorem~\ref{main} reduces the exit problem for (\ref{diffusion}) to the 
minimization problem
\be\lbl{min}
U_\gamma(z)\rightarrow \min, \;\;z\in\partial D.
\ee
In the remainder of this section, we study (\ref{min}) for the weak, strong, and intermediate
coupling strength.

\subsection{The weak coupling regime}\lbl{weak}
In this subsection, we study the minima of $U_\gamma(z)$ on $\partial D$
for small $|\gamma|$. First, we locate the points of minima
of the $U_\gamma(z)$ for $\gamma=0$  (cf. Lemma~\ref{U_0}). Then,
using the Implicit Function Theorem, we  continue them for small $|\gamma|$ 
(cf. Theorem~\ref{imf}).

\begin{lem}\lbl{U_0}
Let $b>0$ in the definition of $D$ (\ref{define-D}) be fixed.
The minimum of $U_0(z)$ on $\p^+ D$ is achieved at $n$ points
\be\lbl{critical}
\xi^i=(\xi^i_1,\xi^i_2,\dots,\xi_n^i),\quad \xi^i_j=\left\{
\begin{array}{cc}
1,&  j=i,\\
-1,& j\neq i,
\end{array}
\right.\;\;j\in [n].
\ee
The minimal value of $U_0(z)$ on $\p D$ is 
\be\lbl{value}
\bar U:=\min_{z\in\p D} U_0(z)={4\over 3}.
\ee
\end{lem}
\pf
Denote
\begin{eqnarray*}
\p_i^+ D &:=& \partial D\bigcap \{z=(z_1,z_2,\dots,z_n)\in\R^n: z_i=1\},\\
\p_i^- D &:=& \partial D\bigcap \{z=(z_1,z_2,\dots,z_n)\in\R^n: z_i=-2-b\},
\end{eqnarray*}
and $\p_i D=\p_i^- D\bigcup \p_i^- D, i\in [n]$.

Consider the restriction of $U_0(z)$ on $\p_1^+ D$
\be\lbl{restrict}
\tilde U_0(y):= U_0((1,y)) ={4\over 3}+\sum_{i=1}^{n-1} F(y_i),
\ee
where $y=(y_1,y_2,\dots, y_{n-1})$.
The gradient of $\tilde U_0$ is equal to 
$$
{\partial\over \p y}\tilde U_0(y) =\mathbf{f}(y):=(f(y_1),f(y_2),\dots, f(y_{n-1}))^\mathsf{T}.
$$
The definition of $f$ (\ref{definef}) implies that 
$U_0$ restricted to $\p_1^+ D$ has a unique critical
point at $z=(1,-\mathbf{1_{n-1}})$ and $U_0((1,-\mathbf{1_{n-1}}))=4/3.$
On the other hand, on the boundary of $\p_1^+ D$, $\p\p_1^+ D$, the minimum
of $U_0(z)$ satisfies
$$
\min_{z\in\p\p_1^+ D} U(z) > {4\over 3},
$$  
for any $b>0$ in (\ref{define-D}).

Likewise, as follows from the definitions of $F$ (\ref{func-1}) and
$D$ (\ref{define-D}), for $z\in \p_1^- D$,
$
U(z) > 4/3,
$
for any choice of $b>0$ in (\ref{define-D}). Thus, $z=(1,-\mathbf{1_{n-1}})$ 
minimizes $U_0$ over $\p_1 D$. The lemma is proved by
repeating the above argument for the remaining faces 
$\p_i D$, 
$i\in [n]\setminus \{1\}$.\\
$\qed$ 

\begin{thm}\lbl{imf}
Suppose $b>0$ in (\ref{define-D}) is sufficiently large.
There exists $\gamma_0>0$ such that for $|\gamma|\le\gamma_0,$
on each face $\partial_i^+ D, i\in [n]$, $U_\gamma(z)$ achieves  minimum
$$
z=\phi^i(\gamma),
$$
where $\phi^i:~[-\gamma_0, \gamma_0]\to\p_i^+ D, i\in [n],$ is a smooth 
function such that 
\be\lbl{phii}
\phi^i(0)=-\mathbf{1_{n-1}},\;\mbox\; {d\over d\gamma}\phi^i(\gamma)\left|_{\gamma=0}\right.=
-l^i,
\ee
and $l^i\in\R^{n-1}$  is the $i$th column of the graph Laplacian $L$ after deleting the
$i$th entry. The equations in (\ref{phii}) are written using the following local coordinates 
for $\p_i^+ D$
$$
(y_1,y_2,\dots,y_{n-1})\mapsto (y_1,y_2,\dots,y_{i-1},1,y_i,\dots,y_{n-2},y_{n-1})\in \p_i^+ D
\subset \R^n.
$$

Moreover, the minimal value of $U_\gamma$ on $\p_i^+ D$ is given by
\be\lbl{valeps}
u^i_\gamma:=\min_{z\in \p_i D} U_\gamma 
={4\over 3}+\gamma~\mathrm{deg}(v_i)
+O(\gamma^2).
\ee
Consequently,
\be\lbl{valeps1}
u_\gamma:=\min_{z\in \p D} U_\gamma
={4\over 3}+\gamma \min_{k\in[n]}~\mathrm{deg}(v_k)
+O(\gamma^2).
\ee
\end{thm}
\pf Let $\tilde U_\gamma(y):=U_\gamma((1,y)),\; y\in\R^{n-1}$  denote the restriction 
of $U_\gamma$ on $\p_1^+ D$:
\be\lbl{compute-restriction}
\tilde U_\gamma (y)={\gamma\over 2}\langle Hz(y), Hz(y)\rangle +\sum_{i=1}^{n-1} F(y_i)
+{4\over 3},
\ee
where
$ 
y=(y_1,y_2,\dots,y_{n-1}),\; z(y):=(1,y_1,y_2,\dots,y_{n-1}).
$

Next, we compute the gradient of $\tilde U_\gamma$:
\be\lbl{grad-1}
{\p \over\p y} \tilde U_\gamma (y)= {\gamma\over 2} {\p \over\p y}\langle Hz(y), Hz(y)\rangle 
-\mathbf{\tilde f}(y),
\ee
where $\mathbf{\tilde f}(y)=(f(y_1), f(y_2),\dots, f(y_{n-1}))$.
Further,
\begin{eqnarray}
{\p \over\p y}\langle Hz(y), Hz(y)\rangle 
\lbl{end-grad}
= 2 \left(
\begin{array}{ccccc}
0 & 1& 0& \dots &0\\
0 & 0& 1& \dots &0\\
\dots  &\dots&\dots& \dots&\dots\\
0 & 0& 0& \dots &1
\end{array}
\right) 
L\left(\begin{array}{c} 1 \\y_1\\y_2\\\dots\\ y_{n-1}
\end{array}\right)=2(L^1y+l^1),
\end{eqnarray}
where $L^i, i\in [n],$ stands for the matrix obtained from $L$ by deleting the 
$i$th row and $i$th column.
By plugging (\ref{end-grad}) in (\ref{grad-1}), we have
\be\lbl{grad}
{\p \over\p y} \tilde U_\gamma (y)=\gamma (L^1y+l^1) -\mathbf{\tilde f}(y).
\ee

The equation for the critical points has the following form
\be\lbl{implicit}
R(\gamma, y):=\gamma (L^1y+l^1) -\mathbf{\tilde f}(y)=0.
\ee
Note that
\be\lbl{verify}
R(0,-\mathbf{1_{n-1}})=0, \quad {\p\over\p\gamma}R(\gamma, y)
\left|_{\gamma=0, y=-\mathbf{1_{n-1}}}\right.= -L^1\mathbf{1_{n-1}}+l^1= 2l^1\neq 0.\footnotemark
\ee   
\footnotetext[1]{
Note that $l^i\neq 0, i\in [n],$ as long as $v_i$ is not an isolated node of $G$.
In particular, if $G$ is connected then $l^i\neq 0\;\forall i\in [n]$.
}
Above we used the relation
\be\lbl{zerosum}
-L^1\mathbf{1_{n-1}}=l^1,
\ee
which follows from the fact that the rows of $L$ sum to zero.

By the Implicit Function Theorem, for small $|\gamma|$,
the unique solution of (\ref{implicit}) is given by 
the smooth function $\phi^1:~[-\gamma_0, \gamma_0]\to \R^{n-1}$
that satisfies
\be\lbl{imfthm}
\phi^1(0)=-\mathbf{1_{n-1}},\; {d\over d\gamma}\phi^1(\gamma)
\left|_{\gamma=0}\right. =
-\left[ {\p\over\p y}R(\gamma, y)
\left|_{\gamma=0, y=-\mathbf{1_{n-1}}}\right.\right]^{-1}
\left[ {\p\over\p\gamma}R(\gamma, y)
\left|_{\gamma=0, y=-\mathbf{1_{n-1}}}\right.\right].
\ee
By taking into account,
$$
{\p\over\p\gamma}R(\gamma, y)
\left|_{\gamma=0, y=-\mathbf{1_{n-1}}}\right.= 2l^1\quad\mbox{and}\quad
{\p\over\p y}R(\gamma, y)\left|_{\gamma=0, y=-\mathbf{1_{n-1}}}\right.
=2I_{n-1},
$$
from (\ref{imfthm}) we have
\be\lbl{derivative}
{d\over d\gamma}\phi^1(\gamma)
\left|_{\gamma=0}\right. = -l^1.
\ee
This shows (\ref{phii}). To show (\ref{valeps}), we use the Taylor expansion
of $\tilde U_\gamma$:
\begin{eqnarray}
\nonumber
\tilde U_\gamma (\phi^1(\gamma))&=& \tilde U_0(-\mathbf{1_{n-1}})+
\gamma \left[ {\p\over\p \gamma}\tilde U_\gamma(\gamma, y)
+\langle {\p\over\p y} \tilde U_\gamma(\gamma, y),
{d\over d\gamma} \phi^1(\gamma)\rangle\right]_{\gamma=0, y=-\mathbf{1_{n-1}}}+
O(\gamma^2)\\
&=& \lbl{taylor}
{2\over 3} +{\gamma\over 2} \langle H (1,-\mathbf{1_{n-1}}),H (1,-\mathbf{1_{n-1}})\rangle +
O(\gamma^2)= {4\over 3} +\gamma~\mathrm{deg}(v_1)+ O(\gamma^2).
\end{eqnarray}

By choosing $b>0$ in (\ref{define-D}) large enough one can ensure that
$\tilde U_\gamma (\phi^1(\gamma))$ for $\gamma\in [-\gamma_0, \gamma_0],$
remains smaller than the values of $U(z)$ on the boundary $z\in\p^+_1 D$.
To complete the proof, one only needs to apply the same argument 
to all other faces of $\p^+ D$, and note that on $\p^- D$, 
the values of $U_\gamma, \gamma\in [-\gamma_0, \gamma_0]$, can be made arbitrarily large by choosing
sufficiently large $b>0$ in (\ref{define-D}).\\
$\qed$
\begin{rem}
The second equation in (\ref{phii}) shows that 
the minima of the potential function lying on the faces corresponding to 
connected cells move towards the common boundaries of these faces, under 
the variation of $\gamma>0$. 
\end{rem}

\subsection{The strong coupling regime}\lbl{strong}
For small $|\gamma|$, the minima of $U_\gamma(z)$ are located near the 
minima of the potential function $\Phi(z)$ (cf. (\ref{func-1})). In this
subsection, we show that for larger $|\gamma|$, the minima of $U_\gamma(z)$
are strongly influenced by the quadratic term $\langle Hz,Hz\rangle $, which corresponds to the 
coupling operator in the differential equation model (\ref{diffusion}).
To study the minimization problem for $|\gamma|\gg 1$,
we rewrite $U_\gamma (z)$ as follows:
\be\lbl{ueps}
U_\gamma (z)=\gamma \left\{ {1\over 2}\langle Hz,Hz\rangle +{1\over\gamma} \Phi(z)\right\}=:
\gamma U^{1\over\gamma}(z).
\ee
Thus, the problem of minimizing $U_\gamma$ for $\gamma\gg 1$ becomes
the minimization problem for
\be\lbl{usup}
U^\lambda(z):=\langle Hz,Hz\rangle+\lambda\Phi (z) \rightarrow \min, \; z\in\partial^+ D, 
\; |\lambda|\ll 1.
\ee

\begin{lem}\lbl{usup0} 
$U^0(z)$ attains the global minimum on $\partial^+ D$
at $z=\mathbf{1_{n}}$:
\be\lbl{u0min}
u^0:=U^0(\mathbf{1_n})=0.
\ee
\end{lem}
\pf
$U^0(z)=\langle Hz,Hz\rangle$ is nonnegative, moreover,
$$
\langle Hz,Hz\rangle=0 \;\mbox{if and only if}\; z\in\ker H=\mathrm{span}\{\mathbf{1_n}\}.
$$
Finally, $\mathbf{1_n}=\ker~H\bigcap \partial^+ D$.\\
$\qed$

\begin{thm}\lbl{small-lambda}
Let $\lambda_1(L^i),\; i\in[n],$ denote the smallest eigenvalue of matrix $L^i$,
obtained from $L$ by deleting the $i$th row and the $i$th column.
Then for $0<\lambda \le 2^{-1}\min_{i\in [n]}\lambda_1(L^i)$,
$U^\lambda(z)$ achieves its minimal value on $\partial D$ at $z=\mathbf{1_n}$:
\be\lbl{u-lambda-min}
u^\lambda:=U^\lambda(\mathbf{1_n})={4\lambda n\over 3},
\ee
provided $b>0$ in the definition of $D$ (cf.~(\ref{define-D})) is sufficiently large. 
\end{thm}
\begin{rem} By the interlacing eigenvalues theorem (cf.~Theorem~4.3.8, \cite{HJ99}),
$
\lambda_1(L^i)\le\lambda_2(L),\;\forall i\in [n].
$
(Note that $L^i$ is not a graph Laplacian.)
Furthermore, $\lambda_1(L^i)>0, \; \forall i\in [n]$, because $G$ is connected 
(cf.~Theorem~6.3, \cite{Biggs}).
With these observations, Theorem~\ref{small-lambda} yields an estimate for
the onset of synchrony in terms of the eigenvalues of $L$:
\be\lbl{onset-syn}
\gamma\ge 2 (\min_{i\in [n]}\lambda_1(L^i) )^{-1}\ge 2(\lambda_2(L))^{-1}.
\ee
Note that (\ref{onset-syn}) yields smaller lower bounds for the onset of
synchrony for graphs with larger algebraic connectivity. In particular,
for the families of expanders $\{G_n\}$ (cf. Example~\ref{ex.5}),
it provides bounds on the coupling strength guaranteeing synchronization
that are uniform in $n\in\N$.   
\end{rem}

For the proof of Theorem~\ref{small-lambda}, we need the following auxiliary 
lemma.
\begin{lem}\lbl{corner}
For $z\in\partial^+D$ there exists $i\in [n]$ such that
\be\lbl{rewrite}
U^\lambda (z)={1\over 2}y^\t L^iy+\lambda\left\{ {4\over 3}+\sum_{j=1}^{n-1} F(1-y_j)\right\},
\ee
where $y=(y_1,y_2,\dots,y_{n-1})^\t$ is defined by
\be\lbl{define-y} 
y_j=\left\{
\begin{array}{cc}
1-z_j, & j\in [i-1], \\
1-z_{j+1}, & j\in [n-1]\setminus [i-1].
\end{array}
\right.
\ee
\end{lem}
\pf
Since $z=(z_1,z_2,\dots, z_n)^\t\in\partial^+ D$, 
$z_i=1$ for some $i\in [n]$. Without loss of generality, let $z_1=1$. Then
$$
z=\left(\begin{array}{c} 1\\
         \mathbf{1_{n-1}}-y\end{array}\right)=: \tilde y, 
\; y=(y_1,y_2, \dots, y_{n-1})^T, \; 0\le y_j\le 3+b,\; j\in [n-1],
$$
and
$$
U^\lambda (z)=\tilde U^\lambda (y):={1\over 2}Q(y)+{4\lambda\over 3} +
\lambda\sum_{j=1}^{n-1} F(1-y_j),
$$ 
where the quadratic function $Q(y)=\langle H\tilde y, H\tilde y\rangle$.
Differentiating $Q(y)$ yields:
$$
{\partial \over \partial y} Q(y)= 
-2\left(
\begin{array}{ccccc}
0 & 1& 0& \dots &0\\
0 & 0& 1& \dots &0\\
\dots  &\dots&\dots& \dots&\dots\\
0 & 0& 0& \dots &1
\end{array}
\right) L \tilde y=
2L^1y\quad \mbox{and}\quad
{\partial^2 \over \partial y^2}Q(y)= 2L^1.
$$
Therefore,
$$
Q(y)=y^\mathsf{T}L^1y.
$$
$\qed$

\pf (Theorem~\ref{small-lambda})
Let $z\in\partial^+D$. By Lemma~\ref{corner}, for some $i\in [n]$
and $y$ defined in (\ref{define-y}), we have
\be\lbl{restate}
U^\lambda (z)={1\over 2}y^\t L^iy+
\lambda\left\{{4\over 3} +\sum_{j=1}^{n-1} F(1-y_j)\right\}.
\ee
We will use the following observations:
\begin{description}
\item[(a)] $L^i$ is a positive definite matrix, and, therefore,
$$
y^\t L^iy\ge \lambda_1(L^i)y^\t y.
$$
\item[(b)] For $\xi\ge 0$,
$$
F(1-\xi)={4\over 3} -\xi^2 +{\xi^3\over 3}\ge {4\over 3} -\xi^2.
$$
\item[(c)] 
$$
\Phi(\mathbf{1_n})= {\lambda 4n\over 3}.
$$
\end{description}
Using (a) and (b), from (\ref{restate}), we have
$$
U^\lambda(z)-U^\lambda(\mathbf{1_n})\ge 
\left(2^{-1}\lambda_1(L^i) -\lambda\right) y^\t y\ge 0
$$
provided $\lambda<2^{-1}\lambda_1(L^i)$.

This shows that $z=\mathbf{1_n}$ minimizes $U_\lambda$ on $\p^+ D$ for 
$\lambda<2^{-1}\min_{i\in [n]}\lambda_1(L^i)$.
On the other hand, on $\p^- D$, $U_\lambda$ can be made arbitrarily large
for any $\lambda>0$ provided $b>0$ in (\ref{define-D}) is sufficiently large.\\  
$\qed$

\subsection{Intermediate coupling strength: formation of clusters}\lbl{inter}
In this subsection, we develop a geometric interpretation of
the spontaneous dynamics of (\ref{1.4}) and (\ref{1.5}).
After introducing certain auxiliary notation, we discuss how the spatial 
location of the minima of $U_\gamma(z)$ on the surface of the $n-$cube 
encodes the most likely activity patterns of (\ref{1.4}) and (\ref{1.5}).
Then we proceed to derive a lower bound on the coupling strength necessary
for the development of coherent structures.

Let $k\in [n],\; 1\le i_1<i_2<\dots<i_k\le n$
and define a $(n-k)$-dimensional face of $D$ as
\be\lbl{face}
\partial^{n-k}_{(i_1,i_2,\dots,i_k)} 
D = \{ (z_1, z_2, \dots, z_n)\in D:~ z_{i_1}=1,\;z_{i_2}=1,\dots,z_{i_k}=1 \}.
\ee
The union of all $(n-k)$-dimensional
faces is denoted by
\be\lbl{union-face}
\partial^{n-k}D = \bigcup_{1\le i_1<i_2<\dots<i_k\le n} \partial^{n-k}_{(i_1,i_2,\dots,i_k)} D.
\ee

Suppose $\xi$ is a point of minimum of $U_\gamma(z)$ on $\partial^+ D$.
If $\xi\in\partial^{n-k}D$, $k\ge 2$, then with high 
probability the network
discharges in $k-$clusters.
The analysis in \S\ref{weak} shows that for small $|\gamma|$ 
there is practically
no correlation between the activity of distinct neurons. 
On the other hand, when the
coupling is strong, all cells fire in unison (cf. \S\ref{strong}). 
Lemma~\ref{cluster} provides a lower bound on the coupling strength
needed for the formation of clusters.

\begin{lem}\lbl{cluster}
Let $\xi\in\partial D$ be a point of global minimum of $U_\gamma$ on 
$\partial D$. If $\xi\in\partial^{(n-k)} D$ for some $k\ge 2$ then
\be\lbl{lower}
\gamma\ge {2\over \max_{j\in[n]}~\mathrm{deg}(v_j)}.
\ee
\end{lem}
\pf Suppose $\hat z\in\partial^{n-2}D$. Without loss of generality,
we assume that 
$$
\hat z=\mathbf{1_n}-(0,0,y_2,y_3,\dots,y_{n-1})=:
\mathbf{1_n}-(0,0,\hat y),\; \hat y\in\R^{n-2}\; y_i\ge 0,\; i\in [n-1]\setminus\{1\}.
$$
Denote $y=(y_1,y_2,y_3,\dots,y_{n-1})=(y_1,\hat y), \; y_1\ge 0,$ 
and $z=\mathbf{1_{n-1}}-(0,y)$.
Thus, 
\begin{eqnarray}\lbl{Uz}
U_\gamma(z) &=& {\gamma\over 2} y^\t L^1y+\sum_{i=1}^{n-1} F(1-y_i) +{4\over 3},\\
\lbl{Uzhat}
U_\gamma(\hat z) &=& {\gamma\over 2} {\hat y}^\t L^{12}\hat y+
\sum_{i=2}^{n-1} F(1-y_i)+{8\over 3},
\end{eqnarray}
where $L^{12}$ is a matrix obtained from $L$ by deleting the first
and the second rows and columns.
The Laplacian of $G$ can be represented as 
$$
L=\mbox{diag}(\mbox{deg}(v_1),\mbox{deg}(v_2),\dots,\mbox{deg}(v_n)) - A,
$$
where the adjacency matrix $A$ is nonnegative (cf. (\ref{def-Lap})). 
Therefore, for nonnegative
$y=(y_1,\hat y)\in\R^{n-1}$,
\be\lbl{ineq1}
y^\t L^1y-{\hat y}^\t L^{12}\hat y \le \mathrm{deg}~(v_2) y^2_2\le \max_{k\in [n]}~\mathrm{deg}~(v_k) y^2_2,
\ee
Further, for any $0<\delta<1$,
\be\lbl{ineq2}
\sum_{i=1}^{n-1} F(1-y_i)-\left\{
\sum_{i=2}^{n-1} F(1-y_i)+{4\over 3}\right\}\le -y_2^2+{1\over 3}y_2^3
\le -(1-\delta) y_2^2,
\ee
provided $0\le y_2< 3\delta$.
The combination of (\ref{Uz})-(\ref{ineq2}) yields
\be\lbl{billy}
U_\gamma(z)\le U_\gamma(\hat z)+ {\gamma\over 2} \max_{k\in [n]}~\mathrm{deg}~(v_k)
-y_2^2+{1\over 3}y_2^3\le U_\gamma(\hat z)+
\left({\gamma\over 2}\max_{k\in [n]}~\mathrm{deg}~(v_k) -1+\delta\right) y^2_2,
\ee
for $y_2\in(0, 3\delta)$. By (\ref{billy}), 
$$
\min_{\hat z\in\partial^{n-2} D} U_\gamma(\hat z)>\min_{z\in\partial^{n-1} D}U_\gamma(z),
$$
provided
\be\lbl{nocluster}
\gamma< {2(1-\delta)\over\max_{k\in [n]}~\mathrm{deg}~(v_k)}.
\ee  
The statement of the lemma follows from the observation above by noting
that $\delta\in (0,1)$ in (\ref{nocluster}) is arbitrary.\\
$\qed$

\section{An alternative view on synchrony} \lbl{another}
\setcounter{equation}{0}
The variational analysis of the previous section  shows that
when the coupling is strong  ($\gamma>(\lambda_2(L))^{-1}$),
the neurons in the network fire together (cf. Theorem~\ref{small-lambda}). 
In this section, we use a complementary approach to studying 
synchrony in the coupled system. We show that strong coupling brings 
about the separation of the timescales in the system's dynamics,
which defines two principal modes in the strong
coupling regime: the fast synchronization and slow 
large-deviation type escape from the potential well.
The analysis of the fast subsystem elucidates  the stability
of the synchronization subspace. In particular, it reveals 
the contribution of the network topology to the robustness of synchrony.
The analysis of the slow subsystem yields the asymptotic
rate of the network activity in the strong coupling regime
(cf. Theorem~\ref{small-lambda}).

\subsection{The slow-fast decomposition}\lbl{slow-fast}
Our first goal is to show that when coupling is strong the 
dynamics of the coupled system (\ref{diffusion}) has two 
disparate timescales. To this end, we introduce the 
following coordinate transformation
$$
z\in\R^n\mapsto (\xi,\eta)\in\R^{n-1}\times\R,
$$
where
\be\lbl{xi-eta}
\xi=\tl Hz\quad \mbox{and} \quad \eta=n^{-1}\mathbf{1_n}^\t z,
\ee
and $\tl H\in \R^{(n-1)\times n}$ is the coboundary matrix corresponding to the spanning
tree $\tl G$ of $G$ (see (\ref{spanning})).

\begin{lem}\lbl{inverse}
Equation (\ref{xi-eta}) defines an invertible linear transformation:
\be\lbl{invert}
z=\eta\mathbf{1_n} +S\xi.
\ee
Matrix $S=(s_{ij})\in\R^{n\times (n-1)}$ satisfies
\be\lbl{Asatisfies}
|s_{ij}|<1, \;\forall (i,j)\in[n]^2 \quad \mbox{and}\quad
\mathbf{1_n}^\t S=0.
\ee
\end{lem}
\pf\; Fix $i\in [n]$. For each $j\in [n]\setminus\{i\}$ there exists
a unique path connecting nodes $v_i\in V(G)$ and $v_j\in V(G)$
and belonging to $\tl G$
\be\lbl{path}
P(i,j)=\sum_{k=1}^{n-1}\sigma_k(i,j) e_k,\quad \sigma_k\in\{0,\pm 1\}.
\ee
Thus,
\be\lbl{zij}
z_j-z_i=\sum_{k=1}^{n-1}\sigma_k(i,j) \xi_k, \;
 j\in [n]\setminus \{i\},
\ee
where $\xi=(\xi_1,\xi_2,\dots,\xi_{n-1})$.
By summing $n-1$ equations (\ref{zij}), adding the identity
$z_i=z_i$ to the resultant equation, and 
dividing the result by $n$, we obtain 
\be\lbl{zi}
z_i=\eta +\sum_{k=1}^{n-1} s_{ik}\xi_k, \;\;\mbox{where}\;\; 
s_{ik}=-n^{-1}\sum_{j=1}^{n-1} \sigma_k(i,j).
\ee
The first inequality in (\ref{Asatisfies}) follows from
the formula for $s_{ij}$ in (\ref{zi}) and $|\sigma_k(i,j)|\le 1$.
To show the second identity in (\ref{Asatisfies}), add up equations
(\ref{zi}) for $i\in[n]$ and use the definition of $\eta$:
$$
n\eta=n\eta +\mathbf{1_n}^\t S\xi \Rightarrow \;(\mathbf{1_n}^\t S\xi=0 \; \forall \xi\in\R^{n-1})\;
\Rightarrow \; \mathbf{1_n}^\t S=0. 
$$
$\qed$

\begin{lem}\lbl{Lhat}
Suppose $G$ is a connected graph and $L\in\R^{n\times n}$ is its Laplacian.
There exists a unique $\hat L\in\R^{(n-1)\times(n-1)}$ such that
\be\lbl{pseudo}
\tl HL=\hat L\tl H,
\ee
where $\tl H\in\R^{(n-1)\times n}$ is the coboundary matrix of the 
$\tl G\subset G$, a spanning tree of $G$. 
The spectrum of $\hat L$ consists of all nonzero eigenvalues of $L$
\be\lbl{spec}
\mathrm{spec} (\hat L)= \mathrm{spec} (L) \setminus \{0\}.
\ee
\end{lem}
\pf\; Since $G$ is connected and $\tl G$ is a spanning tree of $G$,
\be\lbl{rank-span}
\mathrm{rank}~\tl H=n-1 \quad \mbox{and}\quad
\mathrm{ker}~\tl H =\mathrm{ker}~L=\mathrm{span}~\{\mathbf{1_n}\}.
\ee
The existence and uniqueness of the solution of the matrix equation (\ref{pseudo}), $\hat L$,
is shown in Lemma~2.3 of \cite{medvedev10b}. Equation (\ref{spec})
follows from Lemma~2.5 of \cite{medvedev10b}.\\
$\qed$

We are now in a position to rewrite (\ref{diffusion}) in terms
of $(\xi,\eta)$.
\begin{lem}
The following system of stochastic differential equations is equivalent in law
to  (\ref{diffusion}):
\begin{eqnarray}\lbl{xi}
\dot \xi &=& -{\gamma\over 2}\hat L\xi +Q_1(\xi,\eta) +\sigma \tl H\dot W,\\
\lbl{eta}
\dot \eta &=& f(\eta)+ Q_2(\xi) + {\sigma\over\sqrt{n}}\dot w,
\end{eqnarray}
where
\be\lbl{Q12}
Q_1(\xi,\eta)= 2\eta\xi +\mathrm{diag}~(\xi_1,\xi_2,\dots,\xi_{n-1}) \tl S\xi,\quad\mbox{and}\quad 
Q_2(\xi)= n^{-1} \xi^\t S^\t S\xi,
\ee
$$
\tl S=\begin{pmatrix}
\row_1(S)+\row_2(S)\\ \row_2(S)+\row_3(S)\\ \dots \\\row_{n-1}(S)+\row_n(S)
\end{pmatrix}.
$$
Throughout this section, $\dot W$ and $\dot w$ denote the white noise processes
in $\R^n$ and $\R$ respectively.
\end{lem}
\pf\;
After multiplying both sides of (\ref{diffusion}) by $\tl H$
and using (\ref{pseudo}), we have
\be\lbl{derivexi}
\dot\xi =-{\gamma\over 2}\hat L\xi +\tl H\mathbf{f}(z) +\sigma\tl H\dot W_t,
\ee
where
$$
\tl H\mathbf{f}(z) = \begin{pmatrix} 
z_2^2-z_1^2 \\ z_3^2-z_2^2\\\dots \\z_n^2-z_{n-1}^2
\end{pmatrix}=
\begin{pmatrix} 
(z_1+z_2)\xi_1\\ (z_2+z_3)\xi_2\\ \dots \\ (z_{n-1}+z_n)\xi_{n-1}
\end{pmatrix}=
\begin{pmatrix}
(2\eta + \left[\row_1(S)+ \row_2(S)\right]\xi)\xi_1\\
(2\eta + \left[\row_2(S)+ \row_3(S)\right]\xi)\xi_2\\
\dots \\
(2\eta + \left[\row_{n-1}(S)+ \row_{n}(S)\right]\xi)\xi_{n-1}.
\end{pmatrix}
$$
This shows (\ref{xi}). By multiplying (\ref{diffusion}) by
$n^{-1}\mathbf{1_n}^\t$, using  
$\mathbf{1_n}\in\mathrm{ker}~L^\t$, and the definition of
$\eta$, we have 
$$
\dot\eta=f(\eta) +n^{-1} \sum_{j=1}^n
\left\{ f(\eta +\row_j(S)\xi)-f(\eta)\right\} +{\sigma\over\sqrt{n}}\dot w_t.
$$
Here, we are using the fact that the distributions of
$
n^{-1}\mathbf{1_n}^\t \dot W
$ and $\sim {1\over \sqrt{n}} \dot w$ coincide.
Using the definition of $f$ (\ref{definef}) and (\ref{Asatisfies}),
we have 
$$
\sum_{j=1}^n
\left\{ f(\eta +\row_j(A)\xi)-f(\eta)\right\}= 2\eta (\mathbf{1_n}^\t S\xi)
+ \tr (S\xi)(S\xi)^\t=\xi^\t S^\t S\xi.
$$
$\qed$

\subsection{Fast dynamics: synchronization}\lbl{fast}
For $\gamma\gg 1$, the system of equations (\ref{xi}) and (\ref{eta})
has two disparate timescales. The stable fixed point $\xi=0$
of the fast subsystem (\ref{xi}) corresponds to the synchronous 
state of (\ref{diffusion})
$$
z_1=z_2=\dots=z_n.
$$
In this section, we analyze the stability of the steady state of the 
fast subsystem. Specifically, we identify the network properties that
determine the rate of convergence of the trajectories of the fast subsystem 
to the steady state and its degree of stability to random perturbations.
These results elucidate the contribution of the network topology to the
synchronization properties of the coupled system (\ref{diffusion}).

By switching to the fast time (\ref{xi}) and (\ref{eta}), we have
\begin{eqnarray}\lbl{X}
\dot X &=& -\hat L X +\delta Q_1(X,Y)+\tl\sigma \tl H\dot W,\\
\lbl{Y}
\dot Y &=& \delta\left( f(Y)+Q_2(X)\right)+{\tl\sigma\over\sqrt n} 
\dot w, \; 
\tl\sigma=\delta^{3/2}\sigma,
\end{eqnarray}
where
\be\lbl{fast-time}
X(s)=\xi(\delta s),\; Y(s)=\eta(\delta s), \; 0<\delta=2\gamma^{-1}\ll 1.
\ee
The leading order approximation of the fast equation (\ref{X}) does
not depend on the slow variable $Y$:
\be\lbl{X0}
\dot{\tl X}=-\hat L\tl X+\tl\sigma\tl H\dot W.
\ee
The solution of (\ref{X0}) with deterministic
initial condition $\tl X(0)=x\in \R^{n-1}$ is a 
Gaussian random process. The mean vector and the 
covariance matrix
functions
\be\lbl{rho-and-e}
m(s):= \E\tl X(s) \quad \mbox{and}\quad 
V(s):=\E \left[(\tl X(s)-m(s))(\tl X(s)-m(s))^\t\right]
\ee
satisfy linear equations
\be\lbl{linear}
\dot m= -\hat L m\quad \mbox{and}\quad 
\dot V = \hat L V + V \hat L+ \tl{\sigma}^2 \tl H\tl{H}^\t
\ee
Recall that $\hat L$ is a positive definite 
matrix, whose smallest eigenvalue $\lambda_1(\hat L)$ is equal to the 
algebraic connectivity of $G$, $\mathfrak{a}=\lambda_2(L)$ 
(see Lemma~\ref{Lhat}). 
By integrating the first equation in (\ref{linear}), we find that 
\be\lbl{expect}
\left|\E \tl{X}(s)\right| = \left|\exp\{-s\hat L\}x\right|\le  
C_1 \exp\{-\mathfrak{a}s\}x\rightarrow 0,\; s\to\infty,
\ee
for some $C_1>0$.
Thus, the trajectories of the fast subsystem converge in mean to the
stable fixed point $\tl X=0$, which corresponds to the synchronization
subspace of (\ref{diffusion}). Moreover, the rate of convergence is
set by the algebraic connectivity $\mathfrak{a}$.

Further,
\be\lbl{second-moment}
\tr V(s)= \E |\tl X(s)-m(s)|^2=\sum_{i=1}^{n-1} \var\tl X(s) +o(1),\; s\gg 1,
\ee
measures the spread of the trajectories around the synchronization subspace.
By integrating the second equation in (\ref{linear}), we have
\begin{eqnarray}
\nonumber
\sigma^{-2}\tr V(s)&=& 
\tr\left[ \int_0^s \exp\{(u-s)\hat L\} \Lambda \exp\{(u-s)\hat L\} \mathrm{d}u\right]\\
\lbl{asymptot}
 &=&\tr\left[\Lambda\int_0^s 
\exp\{-2\hat L\}u\mathrm{d}u \right] \rightarrow 
{1\over 2}\kappa(G,\tl G), s\to\infty,
\end{eqnarray}
where
\be\lbl{kappa}
\kappa(G,\tl G):=\tr\{\hat{L}^{-1}\Lambda\} \quad\mbox{and}\quad\Lambda=\tl H\tl{H}^\t.
\ee
Parameter $\kappa(G,\tl G)$ quantifies the mean square stability of the synchronization subspace.
In \S\ref{topology}, we show that $\kappa(G,\tl G)$ depends on the properties of the cycle subspace
of $G$.

For small $\tl\sigma>0$, a typical trajectory of (\ref{X0}) converges to a small neighborhood 
of the stable equilibrium at the origin. However, eventually it leaves the neighborhood
of the origin under persistent random perturbations. Next, we estimate the time that the trajectory
of the fast subsystem spends near the origin.

Let $\rho>0$ and
$B_\rho=\{X\in\R^{n-1}:\; |X|<\rho\},\; \rho>0$. By $\tl X(t)$ we denote the solution of (\ref{X0})
satisfying initial condition $\tl X(0)=x\in B_\rho$. Define the first exit time 
of the trajectory of (\ref{X0}) from $B_\rho$: 
\be\lbl{exit}
\tau(\tl X, \rho)=\inf_{t>0}\{ \tl X(t)=\rho\}.
\ee
Using the large deviation estimates (cf. \cite{FW}), we have 
\be\lbl{large-dev}
\lim_{\tl\sigma\to 0}\P_x \left\{\exp\{\tl{\sigma}^{-2} (V_0(\rho)-h)\}
<\tau(x,\rho)< \exp\{\tl{\sigma}^{-2} (V_0(\rho)+h)\}\right\}=1,\; \forall h>0,
\ee
where
\be\lbl{V0}
V_0(\rho)=\min_{|x|=\rho}{1\over 2} \langle Lx, x\rangle ={\mathfrak{a}\rho^2\over 2}
\ee
is the minimum of the positive definite quadratic form $2^{-1} \langle Lx, x\rangle$
on the boundary of $B_\rho$.

By combining (\ref{large-dev}) and (\ref{V0}) we obtain the 
logarithmic asymptotics of the first exit time from $B_\rho$
\be\lbl{log-X}
\tau(\tl X,\rho)\asymp 
\exp\left\{ {\mathfrak{a}\rho^2\over \tl\sigma^2}\right\}.
\ee
For small $\delta>0$ (i.e., for large $\gamma\gg 1$),
(\ref{log-X}) applies to the fast equation (\ref{X}). 
Switching back to the original time, we rewrite the 
estimate for the first exit time for the fast equation
(\ref{xi}):
\be\lbl{log-XX}
\tau (\xi,\rho)\asymp 
\exp\left\{ {\mathfrak{a}\gamma^3\rho^2\over 8\sigma^2}\right\}.
\ee
Finally, choosing $\rho:=\gamma^{-3+\iota\over 2}$ for an arbitrary fixed
$0<\iota<{3\over 2}$, we have 
\be\lbl{log-final}
\tau (\xi,\rho)\asymp 
\exp\{O(\sigma^{-2}\gamma^{\iota})\},\quad \gamma\gg 1,\; 0<\sigma\ll 1.
\ee

\subsection{The slow dynamics: escape from the potential well}
We recap the results of the analysis of the fast subsystem. The trajectories
of the fast subsystem (\ref{xi}) enter an $O(\gamma^{-3+\iota\over 2})$ neighborhood
of the stable equilibrium $\xi=0$  (corresponding to the synchronization subspace 
of the full system) in time $O(\gamma^{-1}\ln\gamma)$ and remain there 
with overwhelming probability over time intervals $O(\exp\{O(\sigma^{-2}\gamma^\iota)\}$.
During this time, the dynamics is driven by the slow equation (\ref{eta}), 
which we analyze next.

While the trajectory of the fast subsystem stays in the $O(\gamma^{-3+\iota\over 2})$ neighborhood
of the equilibrium, the quadratic term in (\ref{eta})
$Q_2(\xi)= O(\gamma^{-3+\iota})$ is small. Thus, the leading order approximation
of the slow equation is independent of $\xi$ on time intervals $O(\exp\{O(\sigma^{-2})\})$
\be\lbl{slow}
\dot{\tilde\eta} = f(\tilde\eta)+ {\sigma\over\sqrt{n}}\dot w.
\ee

Equation (\ref{slow}) has a stable fixed point at the origin, where 
the potential function $F(\eta)$ attains its minimum value (see (\ref{func-1})). 
The escape of the trajectories of (\ref{slow}) from the potential well, defined by 
$F(\eta)$ corresponds to spontaneous synchronized discharge of the coupled subsystem.
Suppose $\tilde \eta(0)=\eta_0<1$ and define
\be\lbl{exit-eta}
\tau(\eta)=\inf\{t>0:~ \eta(t)=1\}.
\ee
By the large deviation theory, we have
\be\lbl{slow-rate}
\tau(\eta)\asymp \exp\left\{ { 2\Delta F n\over \sigma^2}\right\},\;\;
\Delta F:=F(1)-F(0)={4\over 3}.
\ee

Equation (\ref{slow-rate}) provides the estimate for the frequency of spontaneous
activity in the strong coupling regime. Note that (\ref{slow-rate}) is consistent
with the estimate in Theorem~\ref{small-lambda} derived using the variational 
argument. Therefore, the analysis in this section yields the dynamical interpretation 
for the minimization problem for (\ref{min}) in the strong coupling regime. 

We summarize the results of the slow-fast analysis.  In the strong coupling regime,
the dynamics splits into two modes: fast synchronization and slow synchronized
fluctuations leading to large deviation type discharges of the entire network.
By analyzing the fast subsystem, we obtained estimates of stability of the
synchronization subspace. 
The analysis of the slow subsystem yields the asymptotic estimate of the firing rate 
in the strong coupling regime.

\subsection{The network topology and synchronization} \lbl{topology}
The stability analysis in \S\ref{fast} provides interesting insights 
into what structural properties of the network are important for
synchronization. In this subsection, we discuss the implications of 
the stability analysis in more detail.

First, we rewrite 
(\ref{expect}) and (\ref{asymptot}), using the original time 
\begin{eqnarray}\lbl{first}
\left|\E\tl{\xi}(t)\right| &\le& C_1 \exp\{-2^{-1}\mathfrak{a}\gamma^{-1}t\}x,\\
\lbl{second}
\E\left| \tl{\xi}(t)\right|^2 &\to & {8\gamma^{-3}\sigma^2} \kappa(G,\tl G),\;
t\to \infty,
\end{eqnarray}
where $\tl\xi(t)=\tl X(\delta^{-1}t)$.
Equation (\ref{first}) shows that the rate of convergence to 
the synchronous state depends on the coupling strength, $\gamma$,
and the algebraic connectivity of the network, $\mathfrak{a}$.
The convergence is faster for stronger coupling and  larger $\mathfrak{a}$.
This, in particular,
implies that the rate of convergence to synchrony in networks on spectral 
expanders (cf. Example~\ref{ex.5}) remains $O(1)$ as the size of the network
grows without bound. In particular, the families of the random graphs
(cf. Example~\ref{ex.4}) have this property.  In contrast,
for many networks with symmetries (cf. Example~\ref{ex.2}) 
the algebraic connectivity $\mathfrak{a}=o(1)$ as $n\to \infty$,
and, therefore, by (\ref{first}), synchronization requires longer
time, if the size of the network grows.
For a more detailed discussion of the synchronization properties 
of the networks on expanders, we refer the interested reader to 
\cite{medvedev10b}. 

Next, we turn to Equation (\ref{second}), which  characterizes the dispersion 
of the trajectories around the synchronization subspace. 
$\E|\tl\xi|^2$ may be viewed as a measure
of robustness of synchrony to noise or, more generally, to constantly acting 
perturbations. By (\ref{second}), the synchrony is more robust
for stronger coupling, because the asymptotic value of 
$\E\left|\tl\xi(t)\right|^2\to 0$ as $\gamma\to\infty$.
The contribution of the network topology to the mean-square
stability of the synchronous state is reflected in $\kappa(G,\tl G)$. 
Trajectories of the 
networks with smaller $\kappa(G,\tl G)$ are more tightly localized around the
synchronization subspace. 

To explain the graph-theoretic interpretation of $\kappa(G,\tl G)$, we review the 
structure of the cycle subspace of $G$. Recall that the edge
set of the spanning tree $\tl G$ consists of the first $n-1$ edges (see (\ref{spanning})):
$$
E(\tl G)=\{ e_1, e_2,\dots, e_{n-1}\}.
$$
If $G$ is not a tree, then
$$
E(G)\setminus E(\tl G)=\{ e_n, e_{n+1},\dots, e_{n+c-1}\},
$$
where $c$ is the corank of $G$. To each edge  $e_{n+k}, k\in [c]$, there corresponds a 
unique cycle $O_k$ of length $|O_k|$, such that it consists of $e_k$ and
the edges from $E(\tl G)$. The following lemma, relates the value of $\kappa(G,\tl G)$
to the properties of the cycles $\{O_k\}_{k=1}^c$.

\begin{lem}\cite{medvedev10b}\lbl{cycles}
Let $G=(V(G), E(G)), \; |V(G)|=n,$ be a connected graph.
\begin{description}
\item[A] If $G$ is a tree then
\be\lbl{n-1}
\kappa(G,\tl G)=n-1.
\ee
\item[B] Otherwise, let $\tl G\subset G$ be a spanning tree of $G$ a
and $\{O_k\}_{k=1}^c$ be the corresponding independent cycles.
\begin{description}
\item[B.1] Denote 
$$
\mu={1\over n-1}\sum_{k=1}^c (|O_k|-1).
$$
Then
\be\lbl{cool}
{1\over 1+\mu} \le {\kappa(G,\tl G)\over n-1}\le 1,
\ee
\item[B.2] If $0<c<n-1$ then 
\be\lbl{c-small}
1-{c\over n-1}\left(1-{1\over M}\right)\le {\kappa(G,\tl G)\over n-1}\le 1,
\ee
where
$$
M=\max_{k\in [c]} \{|O_k| +\sum_{l\neq k} |O_k\cap O_l|\}.
$$
\item[B.3]
If $O_k, k\in [c]$ are disjoint.
Then 
\be\lbl{disjoint}
{\kappa(G,\tl G)\over n-1}=
1-{c\over n-1}\left(1-{1\over c}\sum_{k=1}^c |O_k|^{-1} \right).
\ee
In particular,
$$
{\kappa(G,\tl G)\over n-1} \le
1-{c\over n-1}\left(1-{1\over \min_{k\in [c]} |O_k|} \right)
$$
and 
$$
{\kappa(G,\tl G)\over n-1}\ge 
1-{c\over n-1}\left(1-{c\over \sum_{k\in [c]} |O_k|} \right)
\ge 
1-{c\over n-1}\left(1-{1\over \max_{k\in [c]} |O_k|} \right).
$$
\end{description}
\end{description}
\end{lem}

The asymptotic estimate of the mean square stability of the synchronization
subspace in (\ref{second}) combined with the estimates of $\kappa(G,\tl G)$
in Lemma~\ref{cycles} show how the structure of the cycle subspace of $G$
translates into the stability of the synchronous state. In \cite{medvedev10b},
one can also find an estimate of the asymptotic stability of the synchronization
subspace in terms of the effective resistance of the graph $G$. These results 
show how the structural properties of the network shape synchronization
properties of the coupled system.

\begin{figure}
\begin{center}
\textbf{a}\includegraphics[height=1.8in,width=2.2in]{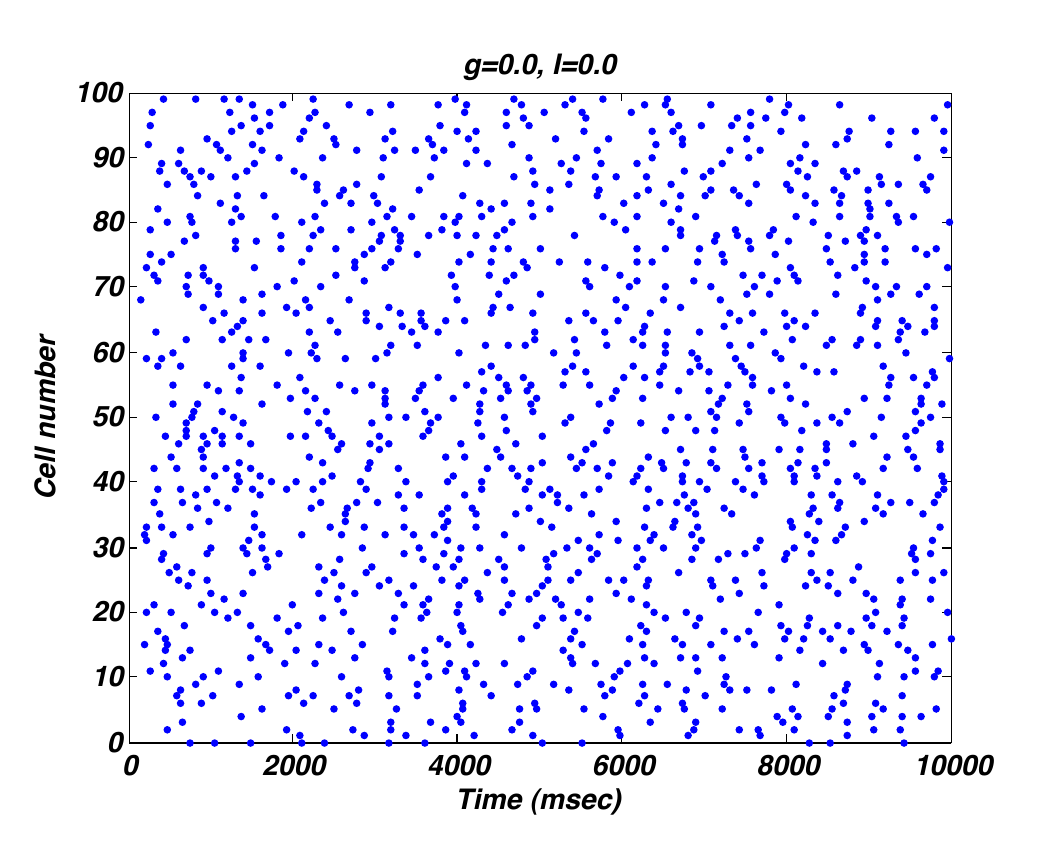}\quad
\textbf{b}\includegraphics[height=1.8in,width=2.2in]{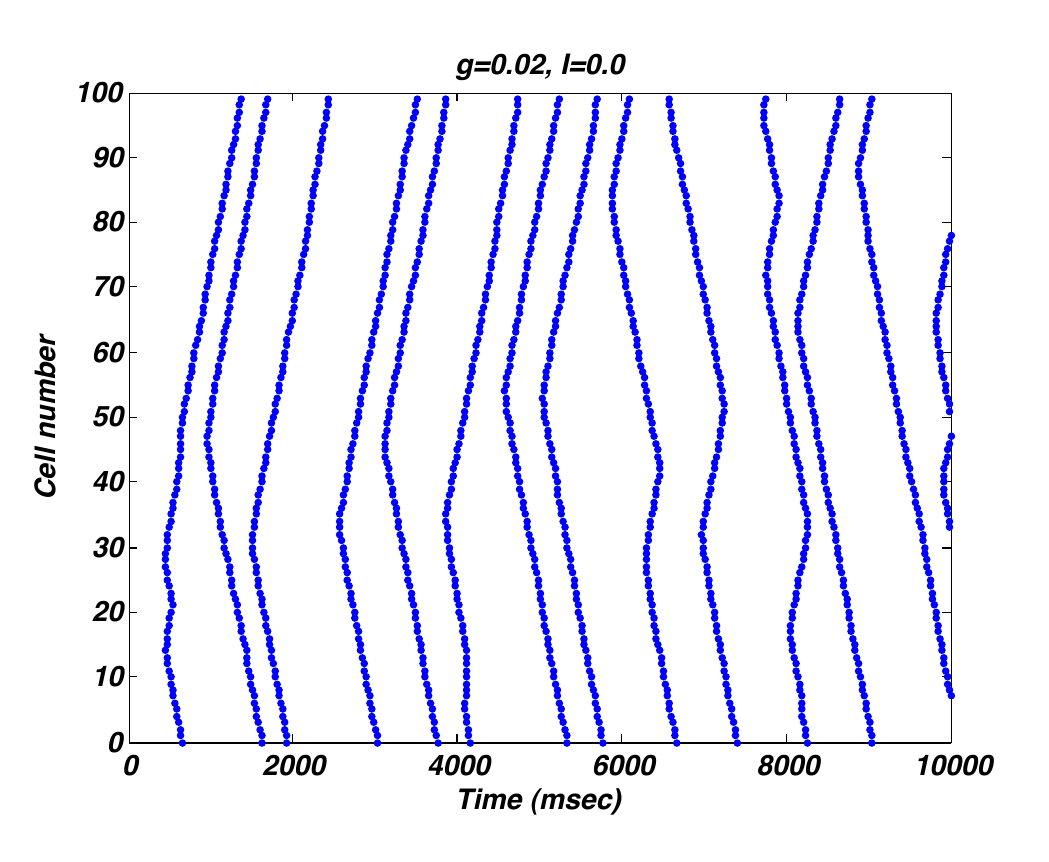} \\
\textbf{c}\includegraphics[height=1.8in,width=2.2in]{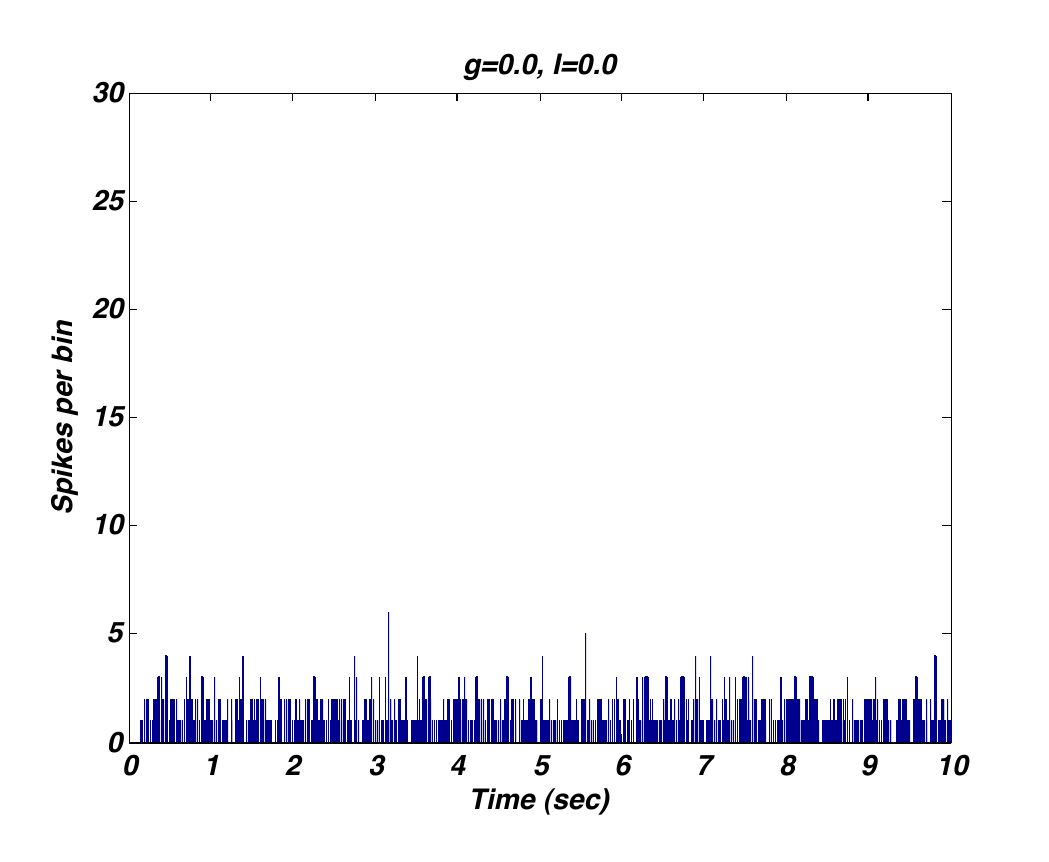}\quad
\textbf{d}\includegraphics[height=1.8in,width=2.2in]{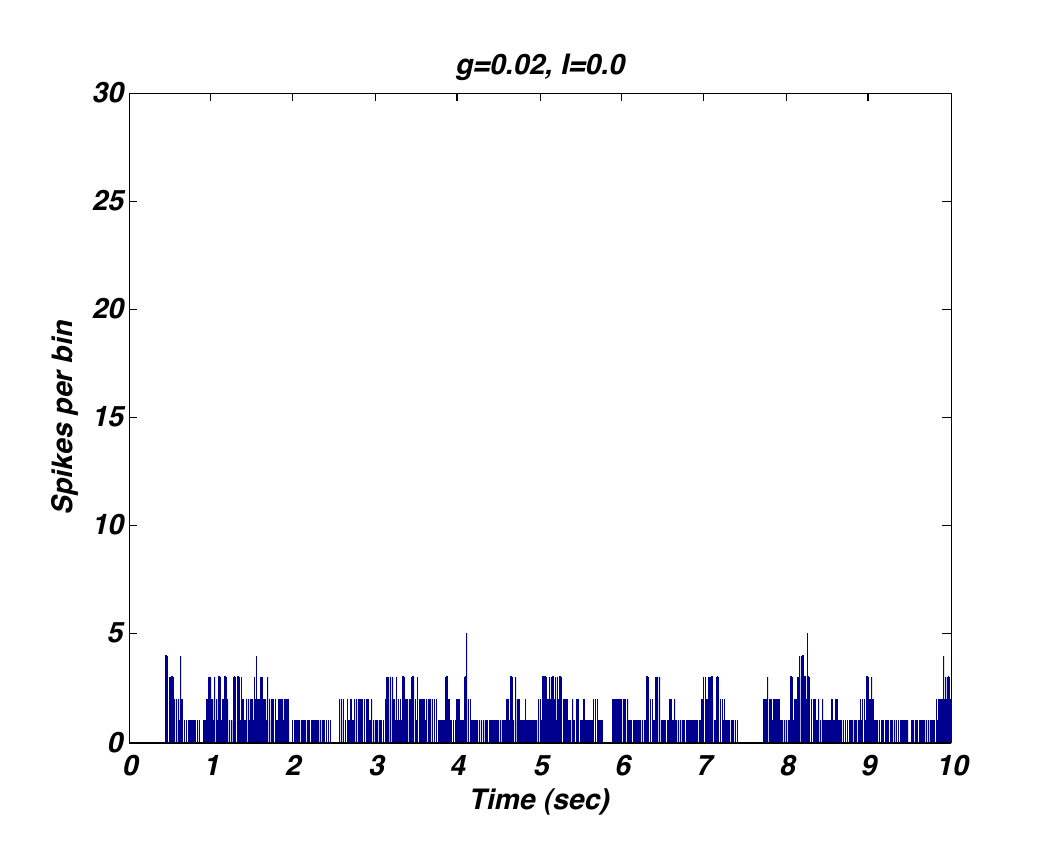}\\
\textbf{e}\includegraphics[height=1.8in,width=2.2in]{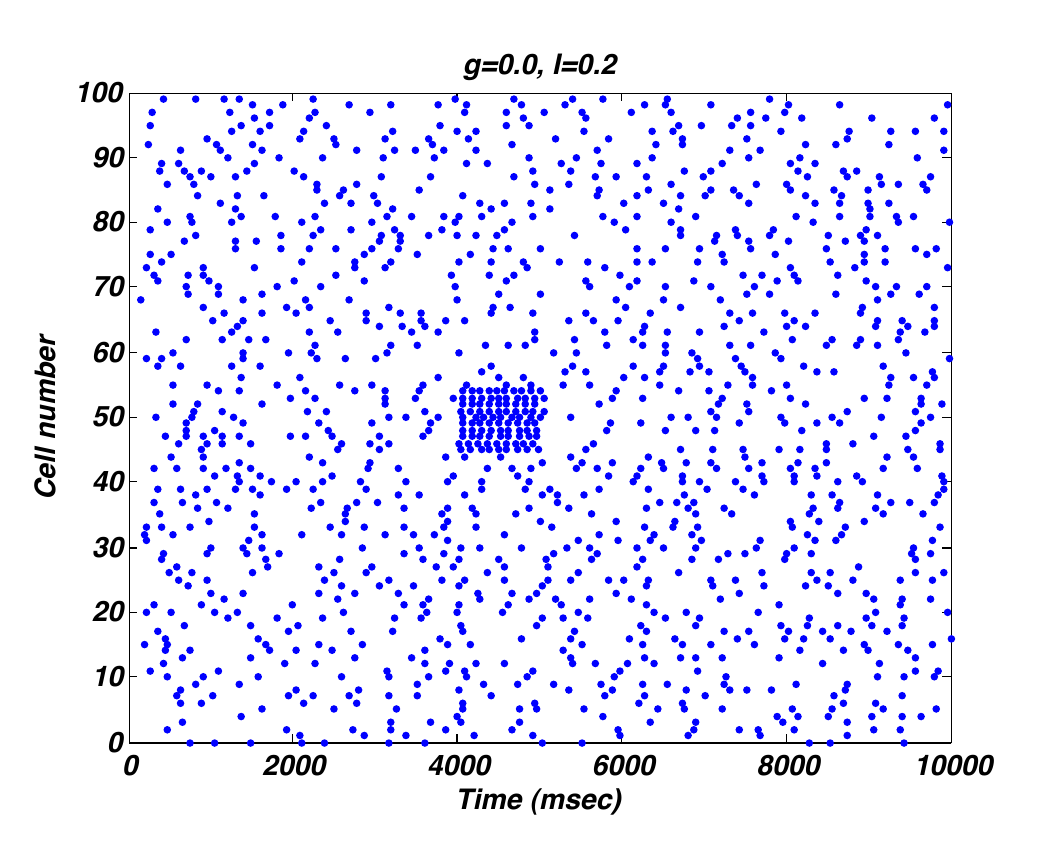}\quad
\textbf{f}\includegraphics[height=1.8in,width=2.2in]{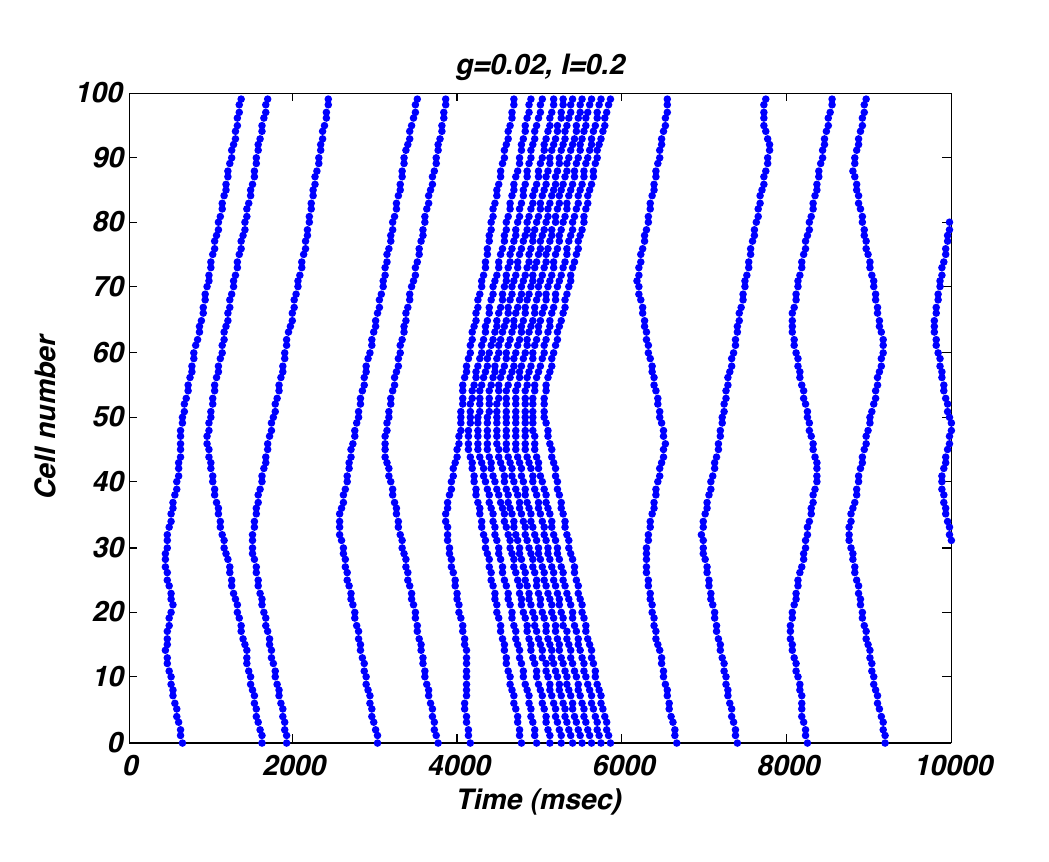}\\
\textbf{g}\includegraphics[height=1.8in,width=2.2in]{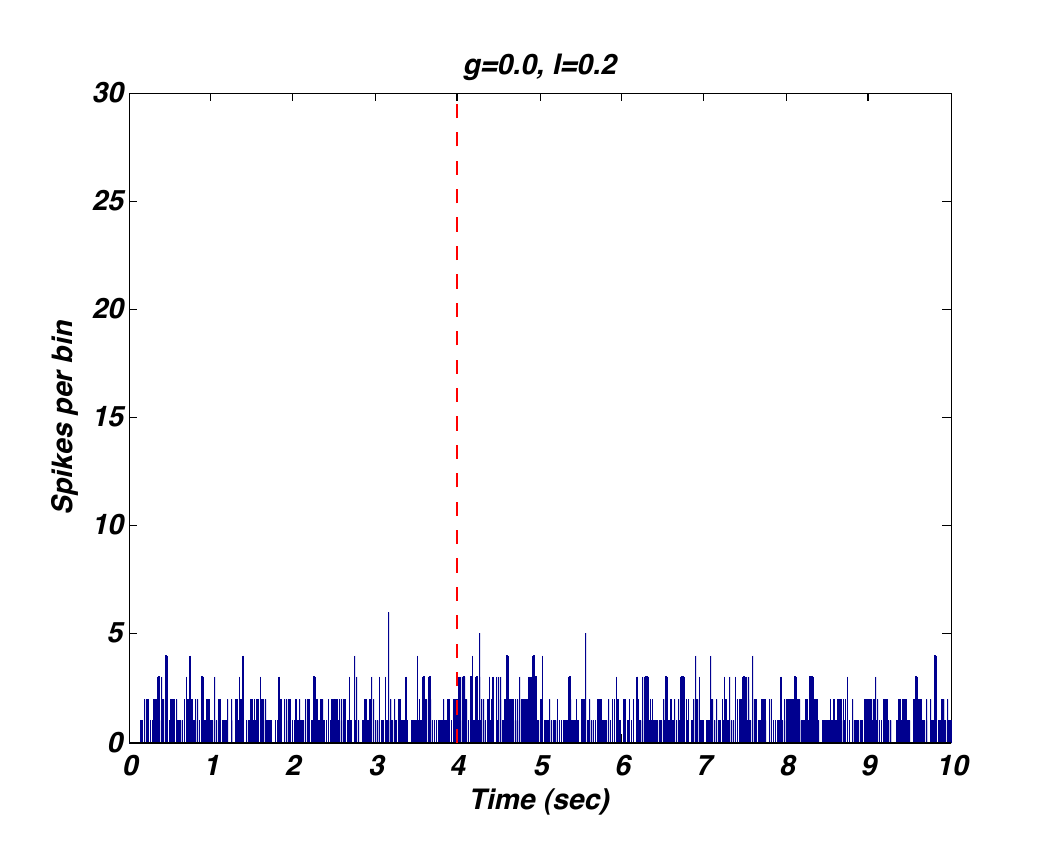}\quad
\textbf{h}\includegraphics[height=1.8in,width=2.2in]{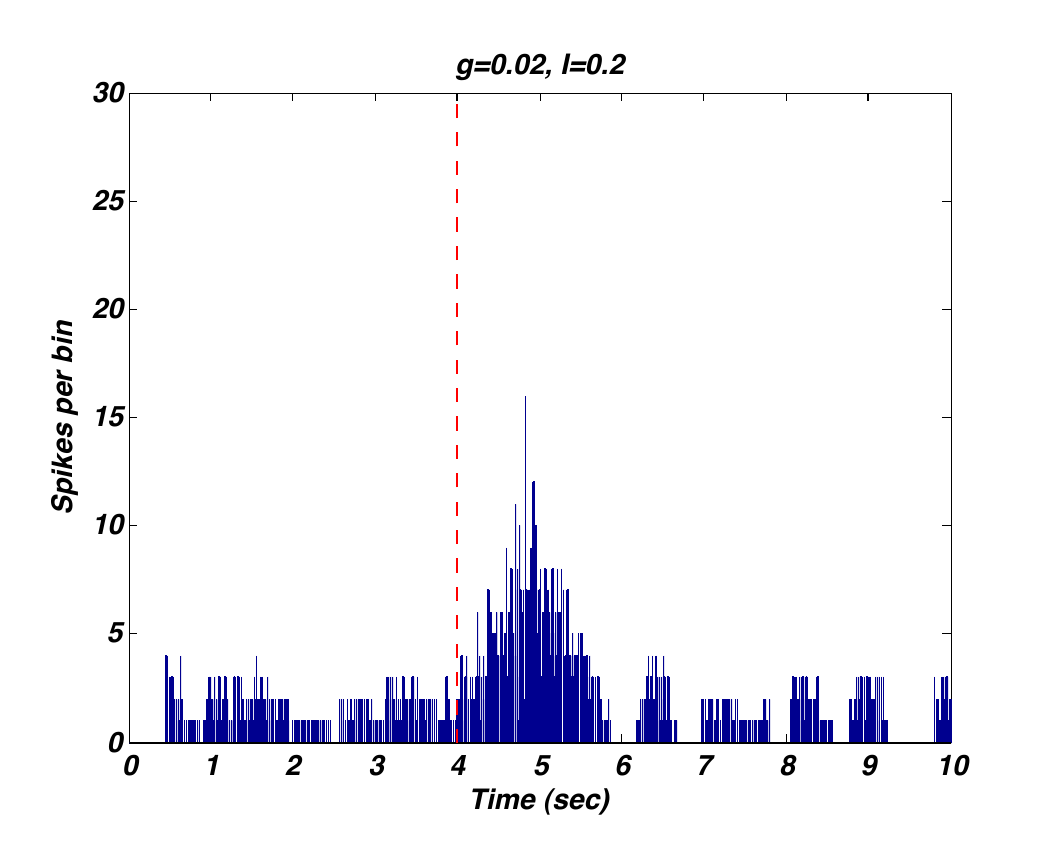}
\end{center}
\caption{The responses to stimulation of the two networks at different values 
of the coupling strength:  $g=0$ (left column) and $g=0.02$ (right column).  
The networks shown
in \textbf{a} and \textbf{b} generate different spatio-temporal patterns.
However, the firing rates corresponding to these activity patterns
are close (see \textbf{c} and \textbf{d}).
When $10$ neurons in the middle of each of these networks receive a 
current pulse, the network responses differ (see \textbf{e} and \textbf{f}).
The firing rate in the first network during the stimulation changes very
little (see \textbf{g}), while the response of the second network
is clearly seen from the firing rate plot (see \textbf{g}). 
}\lbl{f.9}
\end{figure}

\section{Discussion}\lbl{discuss}
\setcounter{equation}{0}
In this paper, we presented a variational method, which reduces the problem 
of pattern-formation in electrically coupled networks of neurons to the minimization
problem for the potential function $U_\gamma(z)$ on the surface of a unit $n-$cube, 
$\partial D$. The variational problem provides geometric interpretation for the
spontaneous dynamics generated by the network. Specifically, 
the location of the
points of minima of the constrained potential function 
$\tilde U_\gamma(z)=U_\gamma(z\in\partial D)$ corresponds to the most likely patterns
of network activity. 

The variational formulation has several important implications for 
the analysis of the dynamical problem. First, the minimization problem has an intrinsic
bifurcation structure. By a bifurcation of the problem (\ref{min}), it is natural
to call the value of the parameter $\gamma=\gamma^*$ corresponding to the structural
changes in the singularity set of $\tilde U_\gamma(z)$. For example, the number
of the minima of $\tilde U_\gamma(z)$ can change due to collisions of the singularities
with each other or with the boundaries of the faces of a given co-dimension 
(a border collision bifurcation). In either case, the qualitative change in the 
configuration of the points of minima of $\tilde U_\gamma(z)$ signals the 
transformation
of the attractor of the randomly perturbed system (\ref{diffusion}).
The study of the constrained minimization problem (\ref{min}) identified
three main regimes
in the network dynamics for weak, intermediate, and strong coupling. These results 
hold for any connected network. We expect that under certain assumptions
on the network topology (e.g., in the presence of symmetries, or alternatively in
networks with random connections), a more detailed description of the bifurcation 
events preceding complete synchronization should be possible.

The analysis of the variational problem has also helped to obtain quantitative 
estimates for the network dynamics. For both weak and strong coupling, we derived the 
asymptotic formulae for the dependence of the firing rate as a function of the coupling 
strength (see (\ref{valeps1}) and (\ref{u-lambda-min})). Surprisingly, in each of these cases, 
the firing rate does not depend on 
the structure of the graph of the network beyond its degree and order.
In particular, the network of equal degree with connectivity patterns 
as different as symmetric and random exhibit the same activity rate 
(see Fig.~\ref{f.4}c).

The geometric interpretation of the spontaneous dynamics yields a novel mechanism
of the formation of clusters. It shows that the network fires in $k-$clusters, 
whenever $U_\gamma(z)$ has a minimum on a co-dimension $k\in[n]$ face of $\partial D$. 
In particular, the network becomes completely synchronized, when the minimum of 
$U_\gamma(z)$ reaches $\mathbf{1_n}\in \partial^+ D\cap \ker (H)$. 
This observation allows one  to estimate the onset of synchronization 
(cf. Theorem~\ref{small-lambda}) and cluster formation (cf. Lemma~\ref{cluster}). 
Furthermore, we show that in the strong coupling regime, the network dynamics has 
two disparate timescales: fast synchronization
is followed by an ultra-slow escape from the potential well. The analysis of the slow-fast 
system yields estimates of stability of the synchronous state in terms of the coupling
strength and structural properties of the network. In particular, it shows the 
contribution of the network topology to the synchronization properties of the network.

We end this paper with a few concluding remarks about the implications of this work 
for the LC network. The analysis of the conductance-based model of the LC network
in this paper agrees with the study of the integrate-and-fire neuron network in 
\cite{UCS99} and confirms that the assumptions of spontaneously active LC 
neurons coupled electrically with a variable coupling strength are consistent
with the experimental observations of the LC network. Following the 
observations in \cite{UCS99} that stronger coupling slows down network
activity, we have studied how the firing rate depends on the coupling strength.
We show that strong coupling results in synchronization and significantly
decreases the firing rate (see also \cite{medvedev09, MZ11}). Surprisingly, 
we found that the rate
can be effectively controlled by the strength of interactions already for very
weak coupling. We show that the dependence of the firing rate on the strength 
of coupling is nonmonotone. This has an important implication for the interpretation
of the experimental data. Because two distinct firing patterns can have similar
firing rates, the firing rate alone does not determine the response of the 
network to external stimulation. This situation is illustrated in Fig.~\ref{f.9}.
We choose parameters such that two networks, the spontaneously active 
(Fig.~\ref{f.9}a) and  the nearly synchronous one (Fig.~\ref{f.9}b), exhibit 
about the same activity rates (see Fig.~\ref{f.9}c,d).    
However, because the activity patterns generated by these networks are 
different, so are their responses
to stimulation (Fig.~\ref{f.9}e,f). The network in the spontaneous firing regime
produces a
barely noticeable response (Fig.~\ref{f.9}g), whereas the response of the synchronized
network is pronounced (Fig.~\ref{f.9}h). Network responses
similar to these were observed experimentally and are associated 
with the good (Fig.~\ref{f.9}h) and poor (Fig.~\ref{f.9}g) cognitive 
performance \cite{UCS99}.
Our analysis suggests that the state of the network 
(i.e., the spatio-temporal dynamics) rather than the firing rate, 
 determines the response of the LC network to afferent stimulation.  

The main hypotheses used in our analysis are that the local dynamical
systems satisfy Assumption~\ref{SN} and interact through electrical coupling.
The latter means that the coupling is realized through one of the 
local variables, 
interpreted as voltage, and is subject to the two Kirchhoff's laws 
for electrical circuits. In this form our assumptions cover many biological, 
physical, and 
technological problems, including power grids, sensor and 
communication networks, 
and consensus protocols for coordination of autonomous agents 
(see \cite{medvedev10b} and references therein). Therefore, the results 
of this work elucidate the principles of pattern formation in 
an important class of problems. 

\noindent {\bf Acknowledgments.} This work was partly supported by 
the NSF Award DMS 1109367 (to GM).

\renewcommand{\theequation}{A.\arabic{equation}}
\section*{Appendix. The parameter values used in the biophysical model 
(\ref{1.1}) and (\ref{1.2}) } 
\setcounter{equation}{0}
\label{sec:A}
To emphasize that the results of this study do not rely on any specific
features of the LC neuron model, in our numerical experiments we used 
the Morris-Lecar model, a common Type I biophysical model of
an excitable cell \cite{RE89}.
This model  is based on the Hodgkin-Huxley paradigm.
The function on the right hand side of the voltage equation (\ref{1.1}), 
$I_{ion}=I_{Ca}+I_{K}+I_l$,
models the combined effect of the calcium and sodium currents, $I_{Ca}$,
the  potassium current, $I_K$, and
a small leak current, $I_l$,
\begin{eqnarray*}
I_{Ca}(v)&=& g_{Ca}m_{\infty}(v)(v-E_{Ca}),\\
I_K(v,n)&=& g_Kn(v-E_K),\\
I_l(v)&=& g_l(v-E_l).
\end{eqnarray*}
Constants $E_{Ca}$, $E_K$, and $E_l$ stand for the reversal potentials and
$g_{Ca}$, $g_K$, and $g_l$ denote the maximal conductances of the corresponding
ionic currents. 
The activation of the calcium and potassium channels are modeled using the 
steady-state functions
$$
m_\infty(v)=0.5\left(1+\tanh\left(\frac{v-\nu_1}{\nu_2}\right)\right)\quad\mbox{and}\quad
n_\infty(v)=0.5\left(1+\tanh\left(\frac{v-\nu_3}{\nu_4}\right)\right),
$$ 
and the voltage-dependent time constant  
$$
\tau(v)=\left(\cosh\left(\frac{v-\nu_3}{2 \nu_4}\right)\right)^{-1}.
$$
The parameter values are summarized the following table.
\begin{center}
\textbf{Table}
\end{center}
\begin{center}
\begin{tabular}{||r|r||r|r||r|r|r|r||r|r||r|r||}
\hline
$E_{Ca}$    & 120$mV$ & $g_K$    &   8$s^{-1}$    & $E_K$    & -84$mV$ & $\nu_2$    &   18$mV$  & $g_{Ca}$    &   4$s^{-1}$ & $\phi$    &  0.067  \\
$g_l$    &   2$s^{-1}$    & $E_l$    & -60$mV$ & $C$    &   20 $\mu F/cm^2$   & $\nu_1$    & -1.2$mV$& $\nu_3$      &   12$mV$ &  $\nu_4$      &  17.4$mV$ \\
\hline
\end{tabular}
\end{center}

\vfill
\break

\end{document}